\documentclass[aps,prX,showpacs,twocolumn,longbibliography,amsmath,amssymb,floatfix]{revtex4-1}
 
\usepackage{graphicx}% Include figure files
\usepackage[normalem]{ulem}
\usepackage{soul}
\usepackage{dcolumn}% Align table columns on decimal point
\usepackage{bm}% bold math
\usepackage{multirow}
\usepackage{esint}
\usepackage{xcolor}
\usepackage{placeins}
\usepackage[utf8]{inputenc}
\usepackage{relsize}
\usepackage{color}
\usepackage{siunitx}
\usepackage[thinlines]{easytable}
\usepackage{float}

\usepackage{ulem}
\usepackage{cancel}

\newcommand{\blue}{\color{black}}

\newcommand{\violet}{\color{black}}

\newcommand{\Chi}{\mbox{\Large$\chi$}}
\newcommand{\Tau}{\mathcal{T}}

\usepackage{color}

\begin{document}

\title{Quantitative comparison of {\blue electrically induced spin and orbital polarizations}
%spin and orbital Hall and Rashba-Edelstein effects
in heavy-metal/$\bm{ 3d}$-metal bilayers}

\author{Leandro Salemi}
 \email{leandro.salemi@physics.uu.se}
\author{Marco Berritta}%
\author{Peter M. Oppeneer}%
\affiliation{Department of Physics and Astronomy, P.\ O.\ Box 516, Uppsala University, SE-75 20 Uppsala, Sweden}

\date{\today}

\begin{abstract}
\noindent
Electrical control of magnetization is of crucial importance for integrated spintronics devices. Spin-orbit torques (SOT) in heavy-metal/ferromagnetic heterostructures have emerged as promising tool to achieve efficiently current-induced magnetization reversal. However, the microscopic origin of the SOT is being debated,with the spin Hall effect (SHE) due to nonlocal spin currents and the spin Rashba-Edelstein effect (SREE) due to local spin polarization at the interface being the primary candidates. We investigate the electrically induced out-of-equilibrium spin and orbital polarizations in pure Pt films and in Pt/$3d$-metal (Co, Ni, Cu) bilayer films using \textit{ab initio} electronic structure methods and linear-response theory. We compute atom-resolved response quantities that allow us to identify the induced spin-polarization contributions that lead to fieldlike SOTs, mostly associated with the SREE, and dampinglike (DL) SOTs, mostly associated with the SHE, and compare their relative magnitude, dependence on the magnetization direction, as well as their Pt-layer thickness dependence. We find that both the FL and DL components contribute to the resulting SOT at the Pt/Co and Pt/Ni interfaces, with the former  contributions being larger at the Pt interface layer and the latter larger in the Co or Ni layers. 
Our calculations show that the electrically-induced transverse orbital polarization is exceedingly larger than the induced spin polarization and present even without spin-orbit coupling, in contrast to the spin polarization.
\end{abstract}

\maketitle

%%%%%%%%%%%%%%%%%%%%%%%%%%%%
%%% INTRODUCTION SECTION %%%
%%%%%%%%%%%%%%%%%%%%%%%%%%%%
%\section{Introduction}

\section{Introduction} 
Electrical control of magnetization has attracted considerable attention because of its potential for high-speed spin-based memories with low-power consumption. Following theoretical predictions  \cite{Slonczewski1996, Berger1996}  it was shown that the magnetization of a ferromagnetic layer in a multilayer stack can be switched with a spin-transfer torque (STT) exerted by a spin-polarized electric current flowing through the magnetic layer in perpendicular direction \cite{Tsoi1998,Myers1999,Wegrowe1999,Katine2000,Brataas2012}.
STT enabled the development of current-operated nonvolatile spin-logic devices, such as the STT-magnetic random access memory (STT-MRAM) \cite{Katine2008}. While STT-based technology is a step forward, there are still shortcomings, such as unintended switching that can occur as the write and read currents flow in the same direction \cite{Jabeur2017}.

A different concept to electrical magnetization switching is the more recently discovered spin-orbit torque (SOT) \cite{Miron2010,Miron2011,Liu2011,Liu2012}. SOT can be observed in a heavy-metal/ferromagnetic bilayer film where the current flows dominantly through the heavy metal and parallel to the ferromagnetic layer. In this configuration, reversible magnetization switching can be achieved in a very energy efficient way and, moreover, have read and write currents flow in distinct directions through the device 
\cite{Kim2013,Fan2013,Safeer2016,Baumgartner2017}.

While it is evident from experiments that the SOT can be used to efficiently reverse the magnetization in the magnetic layer, 
its microscopic origin is still to be fully understood. Two candidates for driving the SOT have attracted much attention: the spin Hall effect (SHE) \cite{Dyakonov1971,Hirsch1999} and the spin Rashba-Edelstein effect (SREE) \cite{Edelstein1990}. Both effects are caused by the spin-orbit interaction, either in the bulk of the material or at an
interface, yet their microscopic appearance is drastically different. The SHE is a nonlocal effect wherein an electrical current generates the flow of a transverse spin current to the boundary of the conducting slab  (see \cite{Kato2004,Wunderlich2005,Hoffmann2013,Sinova2015}) where it exerts a torque on the adjacent ferromagnetic layer. 
The SREE conversely is a local effect: as pointed out by Edelstein \cite{Edelstein1990}, a nonequilibrium spin polarization is generated at a symmetry-broken interface by an electric current  in the presence of Rashba spin-orbit coupling (SOC) \cite{Bychkov1984}.
Both effects have been discussed in the context of SOT switching,  in some cases the SHE was considered as the dominant effect \cite{Liu2011,Liu2012} whereas in other cases the focus was on the SREE \cite{Miron2010,Miron2011,Kim2013,Ciccarelli2016}. In heavy-metal/ferromagnetic bilayer structures both effects are expected to be present simultaneously and will contribute to the fieldlike (FL) SOT and dampinglike (DL) SOT \cite{Freimuth2014,Amin2016,Wimmer2016,Mahfouzi2018,Berger2018,Belashchenko2019,Mahfouzi2020}, yet their relative contribution remains disputed  and continues to be a topic of contemporary investigations \cite{Fan2014Peter,Du2020,Zhu2020Peter}
(see also \cite{Manchon2019} for a recent review). 

First-principles calculations can provide insight in their detailed microscopic origin and offer a way to make a quantitatively comparison \cite{Freimuth2014,Tokatly2015,Wang2016,Amin2018,Belashchenko2019,Mahfouzi2020}.

The SHE and SREE are however not the only magnetic effects that can occur. It was discovered theoretically that, in addition to the spin polarization induced by a current through the SHE, also a nonequilibrium orbital polarization can be induced, which represents an orbital Hall effect (OHE) \cite{Guo2005,Tanaka2008,Kontani2009,Go2018,Go2020}. Similarly, the presence of spatial symmetry breaking in a material was recently shown to lead to a {\blue significant} local orbital polarization, i.e., an orbital Rashba-Edelstein effect (OREE) \cite{Salemi2019}. Both the OHE and OREE are currently only poorly understood, in terms of their relative magnitudes as well as directions of the induced orbital torques. So far several first-principles calculations have been reported for the OHE \cite{Guo2005,Tanaka2008,Kontani2009,Go2018,Go2020}. A direct observation of the induced orbital polarization is yet to be achieved in experiments (see Refs.\ \cite{Stamm2019,Xiao2020} for recent studies).

In this work, we employ relativistic density functional theory (DFT) and Kubo linear-response theory to compute the spin and orbital response to an external electric field for realistic metallic bilayer structures in which Pt is chosen as the heavy-metal material. Specifically, four different systems are investigated: a pure Pt system and three Pt/$3d$-metal bilayer systems, where the $3d$ element is Ni, Co or Cu. For these we compute the spin and orbital conductivity and magneto-electric (ME) tensors resolved for the individual atomic layers in the metallic heterostructures. 

In the following, we first introduce the theoretical framework of linear response within DFT and subsequently apply our formalism to compute the spin and orbital responses for the considered bilayer systems, for various Pt thicknesses. We analyze the spatial symmetry of the spin response, which is embodied in the spin ME susceptibility tensor $\boldsymbol{\chi}^s$. {\blue We show that the direction of the induced spin magnetization $\delta\bm{S}$} depends on the relative directions of the applied electric field $\bm{E}$, the equilibrium magnetization direction $\bm{M}$, and the system geometry. The tensors can be decomposed into odd-in-$\bm{M}$ and even-in-$\bm{M}$ components. {\blue Based on this parity with respect to $\bm{M}$,  we discuss their relationship to the DL and FL components of the SOT.}

The relative importance of those tensor contributions strongly depends on the position of the atomic layer in the slab, and, to a lesser extent, to the thickness of the Pt slab. We investigate furthermore the magnetization-direction dependence of the spin {\blue and orbital} responses. {\blue Whereas the spin response vanishes when the spin-orbit interaction is turned off, off-diagonal components of the orbital ME tensor remain.} 
We compute effective spin-orbit torques on the magnetic Ni and Co layers and compare our results with previously reported values. {\blue Finally, we compare the relative importance of the FL and DL components of $\boldsymbol{\chi}^S$ and discuss their relationship to the SHE ad SREE. We find that the induced spin polarization on the Pt side of the Pt/Ni and Pt/Co interfaces mainly leads to a FL SOT while the induced spin polarization further into the FM layer contributes more to the DL torque component.}

\section{Theory}
\subsection{Linear response}
The materials are modeled within DFT by the relativistic Kohn-Sham Hamiltonian as implemented in WIEN2k \cite{Blaha2018},
\begin{equation}
\hat{H}_0 |n\bm{k}\rangle = \epsilon_{n\bm{k}}|n\bm{k}\rangle
\end{equation}
where $\hat{H}_0$ is the relativistic Kohn-Sham Hamiltonian, $|n\bm{k}\rangle$ the single-electron Kohn-Sham state for band $n$ at wavevector $\bm{k}$ and $\epsilon_{n\bm{k}}$ the corresponding eigenenergy. Under the influence of an external perturbation $\hat{V} = -e\, \hat{\bm{r}} \cdot \bm{E}$ where $e$ is the electron charge, $\bm{E}$ the external electric field and $\hat{\bm{r}}$ the position operator, the %$i^\text{th}$ component ($i=x,y,z$) of $\delta \bm{A}$, the 
change $\delta \bm{A}$ in expectation value of a vectorial observable $\boldsymbol{\mathcal{A}}$ associated to vector operator $\hat{\bm{A}}$, can be expressed within the linear-response formalism \cite{Guo2008,Freimuth2014,Wimmer2016,Mahfouzi2018} as
\begin{equation}
\delta A_i = \sum_{j=x,y,z} \chi^{A}_{ij} ~ E_j.
\end{equation}

The response $\chi^{A}_{ij}$ is expressed in terms of solutions of $\hat{H}_0$,
\begin{equation}
\label{eq:LinResp}
%\small
\begin{split}
\chi^{A}_{ij} &= -\frac{ie}{m_e} \int_{\Omega} \frac{d\bm{k}}{\Omega}
\sum_{n\neq m} \frac{f_{n\bm{k}} - f_{m\bm{k}} }{\hbar \omega_{nm\bm{k}}}~
\frac{A_{mn\bm{k}}^{i} ~ p^j_{nm\bm{k}} }{-\omega_{nm\bm{k}} + i\tau_{\text{inter}}^{-1}}\\
& ~~~ -\frac{ie}{m_e} \int_{\Omega} \frac{d\bm{k}}{\Omega}
\sum_{n} \frac{\partial f_{n\bm{k}} }{\partial \epsilon}~
\frac{A_{nn\bm{k}}^{i} ~ p^j_{nn\bm{k}} }{i\tau_{\text{intra}}^{-1}} \, .
\end{split}
\end{equation}
with $m_e$ the mass of the electron, $f_{n\bm{k}}$ the occupation of Kohn-Sham state $|n\bm{k}\rangle$, $\Omega$ the Brillouin-zone volume, $p^j_{nm\bm{k}}$ the $\hat{p}_j$ momentum-operator matrix element, $A^i_{mn\bm{k}}$ the $\hat{A}_i$-operator matrix element and $ \hbar \omega_{nm\bm{k}} = \epsilon_{n\bm{k}} - \epsilon_{m\bm{k}}$,
the difference of Kohn-Sham eigenergies. As discussed below, we use for $\hat{\bm{A}}$ the spin and orbital angular momentum operators, $\hat{\bm{S}}$ and $\hat{\bm{L}}$, as well as the   the spin and orbital current-density operators, $\hat{\bm{J}}^{\bm{S}}$ and $\hat{\bm{J}}^{\bm{L}} $. {\blue The product $J^{i\bm{S}}_{mn\bm{k}}p^j_{nm\bm{k}}$ is proportional to the spin Berry connection \cite{Xiao2020}.}
The quantity $\tau_{\text{inter}}$ ($\tau_{\text{intra}}$) is the electronic lifetime for inter (intra) band transitions. In {\blue the main part of this work,} $\tau_{\text{inter}}$ and $\tau_{\text{intra}}$ are set to $\hbar \tau_{\text{inter}}^{-1} =$ 0.272 eV and $\hbar \tau_{\text{intra}}^{-1} = 0.220$ eV. Those values have been determined by comparing linear-response calculations to experimental conductivity data for Pt thin films \citep{Stamm2017}. {\blue The influence of the electronic lifetime on the computed nonequilibrium spin and orbital polarizations is discussed in Sec.\ \ref{Sec:lifetime-dep}.}

\subsection{Angular momentum and flow of angular momentum}
\label{secIIB}
The induced angular momentum is composed of a spin and orbital contribution. Let us first focus on the spin part.

The spin operator $\hat{\bm{S}}$ and spin-density current operator $\hat{\bm{J}}^{S_k}$ can be defined as
\begin{align}
\label{eq:Spin_Operator_Definition}
&\hat{\bm{S}} = \frac{\hbar}{2} \Big(\hat{\sigma}_x, \hat{\sigma}_y, \hat{\sigma}_z\Big) ,\\
\label{eq:Spin_Current_Operator_Definition}
&\hat{\bm{J}}^{S_k} = \frac{ \{\hat{S}_k, \hat{\bm{p}}\}}{2 m_e V} ,
\end{align}
where $\hat{\sigma}_x$, $\hat{\sigma}_y$, and $\hat{\sigma}_z$ are the Pauli matrices, $\{.\,,.\}$ denotes the anti-commutator, 
%$\hat{\bm{p}} = (\hat{p}_x, \hat{p}_y, \hat{p}_z)$ the momentum operator, 
$V$ is a reference volume and $k$ ($k=x,y,z$) an index specifying the direction of the spin polarization carried by the spin-current density. In this work, $V$ refers to the individual atomic spheres, allowing us to compute atom-projected quantities (see Appendix A for details).

Using the linear-response formalism, we can compute the out-of-equilibrium electrically induced spin angular momentum $\delta \bm{S}$ as well as the %electrically 
induced spin-current density $\bm{J}^{S_k}$, using
\begin{align}
\label{eq:Spin_SpinSusceptibility_Efield}
\delta \bm{S} &= \boldsymbol{\chi}^S ~  \bm{E} \, ,\\
\label{eq:SpinCurrent_SpinConductivity_Efield}
\bm{J}^{S_k}& = \boldsymbol{\sigma}^{S_k} ~ \bm{E} \, ,
\end{align}where $\boldsymbol{\chi}^S$ is the spin ME susceptibility tensor and $\boldsymbol{\sigma}^{S_k}$ the spin conductivity tensor. 
Both $\boldsymbol{\chi}^S$ and $\boldsymbol{\sigma}^{S_k}$ are real $2^{\text{nd}}$-rank tensors, but note that, due to the spin component dependence, the spin conductivity tensor can be associated with a $3^{\rm rd}$-rank tensor, $\boldsymbol{\sigma}^{\bm{S}}$.

Analogous quantities can be straightforwardly defined for the orbital angular momentum $\hat{\bm{L}}$. Thus, we can define the orbital ME susceptibility tensor $\boldsymbol{\chi}^L$ and orbital conductivity tensor $\boldsymbol{\sigma}^{L_k}$,
\begin{align}
\delta \bm{L} &= \boldsymbol{\chi}^L ~  \bm{E}\, ,\\
\bm{J}^{L_k} &= \boldsymbol{\sigma}^{L_k} ~ \bm{E} \, ,
\end{align}
where $\delta \bm{L}$ is the out-of-equilibrium electrically induced orbital angular momentum and $\bm{J}^{L_k}$ the induced orbital current density.

\subsection{Computational methodology}
\label{CompMeth}
\begin{figure}[t!]
\centering
\includegraphics[width=\linewidth]{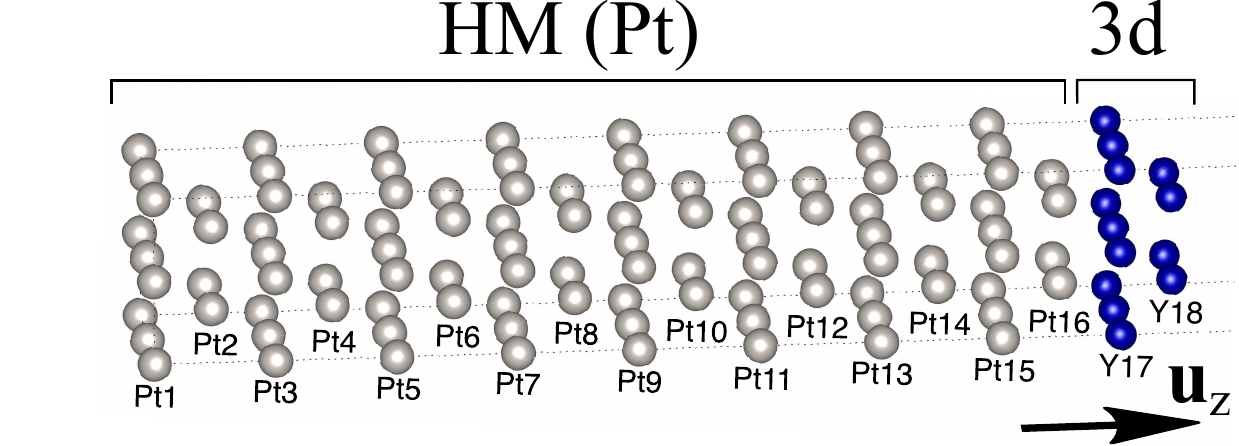}
\caption{Schematic of a typical system studied in this work, a $n$Pt/2$Y$ bilayer. There are $n$ ($=16$, here) monolayers of Pt heavy metal (HM) capped by two $Y$ monolayers, where $Y$ is Ni, Co, or nonmagnetic Cu or Pt. The $z$ axis is taken normal to the slab, with unit vector $\bm{u}_z$. Each atomic plane is numbered with an index, where index $1$ refers to the Pt atomic-layer interfaced with vacuum,  $n$ to the Pt atomic-layer interfaced with the 3$d$ element in layer $n+1$, and $n+2$ labels the top layer  at the vacuum interface.}
\label{fig:16Pt2Y_Schematic}
\end{figure}

The bilayer structures that are studied here consist of several Pt monoatomic layers that are covered with two monoatomic layers of the $3d$ elements Ni, Co or Cu (see Fig.\ \ref{fig:16Pt2Y_Schematic}). For comparison, we also study the pure Pt system, where the top two monolayers consist of Pt. The nomenclature used in this paper is the following: we denote our systems by $n$Pt/2$Y$ where $n$ is the total number of Pt monolayers and $Y$ is either Ni, Co, Cu or Pt. The minimum total number of Pt monolayers used in our calculations is 2 while the maximum is 18 (denoted as 16Pt/2Pt). The maximum thickness achieved is then $\sim$3.2 nm. The direction normal to the interfaces is taken as the $z$ axis.

The monoatomic layers are labeled from $z=1$ for the Pt monoatomic layer at the interface with vacuum (leftmost layer in Fig.\ \ref{fig:16Pt2Y_Schematic}) to $z=n+2$ for the $Y$ monoatomic layer at the interface with vacuum (rightmost layer in Fig.\ \ref{fig:16Pt2Y_Schematic}). Particular positions can be identified, like $z=n$ for the Pt monoatomic layer at the Pt/$Y$ interface and $z=n+1$ for the $Y$ monoatomic layer at the Pt/$Y$ interface.

To compute the spin and orbital susceptibility and conductivity tensors, we use the following three-step procedure.
\begin{enumerate}
\itemsep0em
\item The cell parameters and atomic positions of the heterostructures are fully relaxed with the DFT package SIESTA \cite{Soler2002}.
\item Using the relaxed atomic positions, the ground-state Kohn-Sham wavefunctions and energies are selfconsistently computed with the accurate full-potential, all-electron DFT package WIEN2k \cite{Blaha2018}.
\item Using the relativistic Kohn-Sham wavefunctions and energies, we compute the response tensors defined by Eq.~(\ref{eq:LinResp}).
\end{enumerate}

As the DFT packages used employ full 3D periodic boundary conditions,  all heterostructures contain 20 {\AA} of vacuum to avoid spurious interactions with neighboring simulation cells. More details on the computational recipe are given in Appendix \ref{ap:Computational_Details}.

\subsection{Symmetry considerations}
{\blue
\subsubsection{Symmetry of the ME tensors}
}
\label{symmetry-ME}
{\blue The structures studied in this work are tetragonal, with nonmagnetic point group 4\emph{mm}. The response tensor $\boldsymbol{\chi}^S$ can be expanded as
\begin{equation}
\label{eq:Magn_Expansion}
\chi^S_{ij} = {\chi}^{(0)S}_{ij} 
+ \sum_k \chi^{(1)S}_{ijk} ~ \mathcal{M}_k 
+ \sum_{kl} \chi^{(2)S}_{ijkl} ~ \mathcal{M}_k ~ \mathcal{M}_l + ...
\end{equation}
where $\mathcal{M}_k$ is the $k$-\text{th} component of the {\blue unit vector of} magnetization, i.e., $\bm{M} = |\bm{M}| \bm{\mathcal{M}}$, with $\bm{M}$ the equilibrium magnetization. As detailed in Ref.~\cite{Zelezny2017}, for $\bm{M} \parallel  \bm{u}_z$, $\boldsymbol{\chi}^S$ can be written as}
\begin{equation}
\boldsymbol{\chi}^S =
\begin{pmatrix}
\chi_{xx}^S & \chi_{xy}^S & 0 \\[0.5em]
\chi_{yx}^S & \chi_{yy}^S  & 0 \\[0.5em]
0 & 0  & \chi_{zz}^S
\end{pmatrix} ~~[ \bm{M} \parallel \bm{u}_z ],
\label{chi_uz}
\end{equation}
where $\chi_{xy}^S = - \chi_{yx}^S$ and $\chi_{xx}^S = \chi_{yy}^S \neq \chi_{zz}^S$. With $\bm{M}$ out-of-plane, the system exhibits an in-plane x/y spatial symmetry, which is fully recovered in our calculations. The $\boldsymbol{\chi}^S$ tensor can be further decomposed into an odd-in-$\bm{M}$ and even-in-$\bm{M}$ component,
\begin{equation}
\boldsymbol{\chi}^S(\bm{M}) = \boldsymbol{\chi}_{o}^S(\bm{M}) + \boldsymbol{\chi}_{e}^S(\bm{M}),
\end{equation}
with specifically,
\begin{subequations}
\begin{align}
\boldsymbol{\chi}_{e}^S &=
\begin{pmatrix}
0 & \chi_{xy}^S & 0 \\
\chi_{yx}^S & 0  & 0 \\
0 & 0  & 0 \\
\end{pmatrix},\\
\boldsymbol{\chi}_{o} &=
\begin{pmatrix}
\chi_{xx}^S & & 0 \\
0 &  \chi_{yy}^S  & 0 \\
0 & 0  & \chi_{zz}^S\\
\end{pmatrix}.
\end{align}
\label{chi-z-symm}
\end{subequations}
A nonzero odd-in-$\bm{M}$ part can obviously not exist for nonmagnetic systems ($n$Pt/$2$Pt and $n$Pt/$2$Cu), which is as well recovered in our calculations.
The spin response is highly dependent on the magnetization direction. Setting the magnetization in plane, $\bm{M} \, || \, \bm{u}_x$, the $\boldsymbol{\chi}^S$ tensor can be written as
\begin{equation}
\boldsymbol{\chi}^S =
\begin{pmatrix}
0 & \chi_{xy}^S & \chi_{xz}^S \\[0.3em]
\chi_{yx}^S & 0  & 0 \\[0.3em]
\chi_{zx}^S & 0  & 0
\end{pmatrix} ~~[ \bm{M} \parallel \bm{u}_x ],
\label{chi-x-symm}
\end{equation}
{\blue being different from} the $\bm{M} \, || \, \bm{u}_z$ case. Now the $\chi_{xy}$, $\chi_{yx}$ elements are even-in-$\bm{M}$ and the $\chi_{xz}$, $\chi_{zx}$ elements odd-in-$\bm{M}$. We can furthermore mention already that the orbital $\boldsymbol{\chi}^L$ tensor has the same nonzero elements with the same $\bm{M}$ parity. {\blue Note that the magnetization direction breaks the $x$/$y$ symmetry such that the $xy$ and $yx$ components of $\chi^{S/L}$ may be different. Such difference, if relevant, comes from terms with powers of $\mathcal{M}$ of order higher than two in  Eq.~(\ref{eq:Magn_Expansion}). As we will see later, those higher order corrections are indeed relevant for the spin response. The symmetry of the system is naturally incorporated in the electronic density $n(\bm{r})$ and magnetization density $\bm{m}(\bm{r})$ computed with DFT. As a result, our calculations, in which we do not impose any symmetry constraints, provide the full $\bm{M}$-dependent response, i.e., the left hand-side of Eq.\ (\ref{eq:Magn_Expansion}).}

Depending on the relative orientation of the induced spin polarization $\delta \bm{S}$ with respect to (1) the applied electric field $\bm{E}$, (2) the normal direction $\bm{u}_z$ and (3) the equilibrium magnetization vector $\bm{M}$, the components of the ME susceptibility $\boldsymbol{\chi}^S$ can be classified according to three categories:
\begin{itemize}
\itemsep0em 
\item $\bm{E}$-transverse components ($\bm{E}_\bot$):
\begin{equation}
\label{eq:Symmetry_Relation_Etrans}
\delta \bm{S} \propto \bm{E} \times \bm{u}_z ,
\end{equation}
\item $\bm{M}$-transverse components ($\bm{M}_\bot$):
\begin{equation}
\label{eq:Symmetry_Relation_Mtrans}
\delta \bm{S} \propto (\bm{E} \times \bm{u}_z) \times \bm{M} ,
\end{equation}
\item $\bm{M}$-longitudinal component ($\bm{M}_\parallel$):
\begin{equation}
\label{eq:Symmetry_Relation_Mlong}
\delta \bm{S} \propto \bm{M} \text{ when } \bm{E} \parallel \bm{u}_z .
\end{equation}
\end{itemize}

{
\blue
\subsubsection{Symmetry of the SOT}
\label{symmetry-SOT}
The SOT $\bm{\mathcal{T}}$ felt by the equilibrium magnetization $\bm{M}$ can be written as
\begin{equation}
\bm{\mathcal{T}}=\bm{M} \times \delta \bm{B},
\end{equation}
where $\delta \bm{B}$ is the SOT effective magnetic field, which originates from the electrically-induced change in the exchange-correlation effective field $\bm{B}_\text{XC}$. We can approximate $\delta \bm{B}$ as
\begin{equation}
\label{eq:effective_sot_field}
\delta \bm{B} \approx
|\bm{B}_\text{XC}| \frac{\delta \bm{S}}{|\bm{S}|} ,
\end{equation}
where $\bm{S}$ is the equilibrium value of the expectation value of $\hat{\bm{S}}$. Essentially, Eq.~(\ref{eq:effective_sot_field}) assumes that the direction of the exchange-correlation effective field changes while its strength is unaffected. Note that because the spatial distribution of $\delta \bm{S}$ may differ for different orbital characters (e.g., $s$, $p$, $d$), Eq.~(\ref{eq:effective_sot_field}) is an estimate of $\delta \bm{B}$.

Using $\bm{M} = -2 \mu_B \bm{S}$, where $\mu_B$ is the Bohr magneton, the torque can be expressed as
\begin{equation}
\bm{\mathcal{T}} =
-2 \mu_B|\bm{B}_\text{XC}| ~ \big(\bm{\mathcal{M}}
 \times  \delta \bm{S}\big)
\end{equation}
where we used that $\bm{\mathcal{M}} = \bm{S}/|\bm{S}|$. With the induced spin polarization  %Injecting
[Eq.\ (\ref{eq:Spin_SpinSusceptibility_Efield})] 
this can be written as
\begin{equation}
\bm{\mathcal{T}} =
-2 \mu_B|\bm{B}_\text{XC}| ~ \Big[\bm{\mathcal{M}}
 \times  \Big( \bm{\chi}^S \bm{E}\Big)\Big] \, .
 \label{eq:SOT-T-expression}
\end{equation}
Since $\bm{\chi}^S$ can be decomposed in odd-in-$\bm{M}$ and even-in-$\bm{M}$ components, we can always decompose the SOT in an odd (o) and an even (e) component, 
\begin{equation}
\bm{\mathcal{T}}
=\bm{\mathcal{T}}_\text{\hspace{-0.1cm}o}
+ \bm{\mathcal{T}}_\text{\hspace{-0.1cm}e}.
\end{equation}
%where the subscript ``o" stands for odd and ``e" for even. 
We can compute $\bm{\mathcal{T}}_\text{\hspace{-0.1cm}o}$ and $\bm{\mathcal{T}}_\text{\hspace{-0.1cm}e}$, e.g.\ for an in-plane field, $\bm{E} = |\bm{E}|~\bm{u}_x$. When $\bm{M} || \bm{u}_z$, using Eq.\ (\ref{chi-z-symm}) we obtain
\begin{subequations}
\begin{align}
\bm{\mathcal{T}}_\text{\hspace{-0.1cm}o} &= ~ + ~
2 \mu_B|\bm{B}_\text{XC}| ~ |\bm{E}| ~
\chi^S_{yx} ~ \bm{u}_x,\\
\bm{\mathcal{T}}_\text{\hspace{-0.1cm}e} &=~ - ~
2 \mu_B|\bm{B}_\text{XC}| ~ |\bm{E}| ~
\chi^S_{xx}~ \bm{u}_y,
\end{align}
\end{subequations}
while for $\bm{M} \, || \, \bm{u}_x$ this yields [see Eq.\  (\ref{chi-x-symm})]
\begin{subequations}
\begin{align}
\bm{\mathcal{T}}_\text{\hspace{-0.1cm}o} &= ~ - ~
2 \mu_B|\bm{B}_\text{XC}| ~ |\bm{E}| ~
\chi^S_{yx} ~ \bm{u}_x,\\
\bm{\mathcal{T}}_\text{\hspace{-0.1cm}e} &=~ + ~
2 \mu_B|\bm{B}_\text{XC}| ~ |\bm{E}| ~
\chi^S_{zx}~ \bm{u}_y.
\end{align}
\end{subequations}

We can now easily establish the link between the symmetry of the components of $\bm{\chi}^S$, their evenness or oddness in $\bm{M}$, and the commonly discussed \textit{fieldlike}  torque $\bm{\mathcal{T}}_\text{\hspace{-0.0cm}FL}$  and 
\emph{dampinglike} torque $\bm{\mathcal{T}}_\text{\hspace{-0.0cm}DL}$,
specifically,
\begin{subequations}
\begin{align}
\label{eq:FL_Torque_General}
\bm{\mathcal{T}}_\text{FL} &\propto \bm{M} \times (\bm{E} \times \bm{u}_z),\\
\label{eq:DL_Torque_General}
\bm{\mathcal{T}}_\text{DL} & \propto \bm{M} \times [\bm{M} \times (\bm{E} \times \bm{u}_z)].
\end{align}
\end{subequations}
We recognize that $\bm{\mathcal{T}}_\text{\hspace{-0.1cm}o}$, and therefore the $\bm{E}_\bot$ components, correspond to $\bm{\mathcal{T}}_\text{\hspace{-0.0cm}FL}$ while $\bm{\mathcal{T}}_\text{\hspace{-0.1cm}e}$, and therefore the $\bm{M}_\bot$ components, correspond to $\bm{\mathcal{T}}_\text{\hspace{-0.0cm}DL}$.}

A similar classification can be carried out for the cases where $\bm{M} \, || \, \bm{u}_x$ {\blue and $\bm{M} \, || \, \bm{u}_y$.} The classification and symmetry relations for these magnetization directions are summarized for convenience in Table \ref{tab:MSymmetry}.

\begin{table}
\normalsize
\centering
\caption{The $\bm{E}$-transverse ($\bm{E}_\bot$), $\bm{M}$-transverse ($\bm{M}_\bot$), and $\bm{M}$-longitudinal ({$\bm{M}_{\parallel}$}) components of the $\boldsymbol{\chi}$ tensor for $\bm{M}$ parallel to $\bm{u}_z$, $ \bm{u}_x $, or $ \bm{u}_y$.
Each row summarizes the equivalency of the $\boldsymbol{\chi}$ components for the three magnetization directions. The spatial symmetry
{\blue of the induced spin polarization $\delta \bm{S}$}
{\blue and whether it leads to a dampinglike or fieldlike spin-orbit torque $\bm{\mathcal{T}}$ are also provided.}}
\label{tab:MSymmetry}
\begin{TAB}(r,0.6cm,1.2cm)[1pt]{|c|c|c|c|c|c|}{|c|c|c|c|}% (rows,min,max)[tabcolsep]{columns}{rows}\hline
 & $\bm{M} \parallel \bm{u}_z$ & $\bm{M} \parallel \bm{u}_x$ & $\bm{M} \parallel \bm{u}_y$ & ~$\delta \bm{S}$ dependence ~ &  \hspace*{0.2cm}$\bm{\mathcal{T}}$ \hspace*{0.2cm} \\
$\hspace{0.1cm}\bm{E}_\bot\hspace{0.1cm}$ & $\Chi_{xy/yx}$ & $\Chi_{xy/yx}$ & $\Chi_{xy/yx}$ & $\bm{E} \times \bm{u}_z$ & FL\hspace{0.2cm}\\
$\bm{M}_\bot$ & $\Chi_{xx/yy}$ & $\Chi_{zx}$ & $\Chi_{zy}$ & $\bm{M} \times (\bm{E} \times \bm{u}_z)$ &  DL\\
$\bm{M}_\parallel$ & $\Chi_{zz}$ & $\Chi_{xz}$ & $\Chi_{yz}$ & $\bm{M}$ ~($\bm{E} \parallel \bm{u}_z$)& $-$ \\
\end{TAB}
\end{table}

\begin{figure*}[ht!]
\centering
\includegraphics[width=\linewidth]{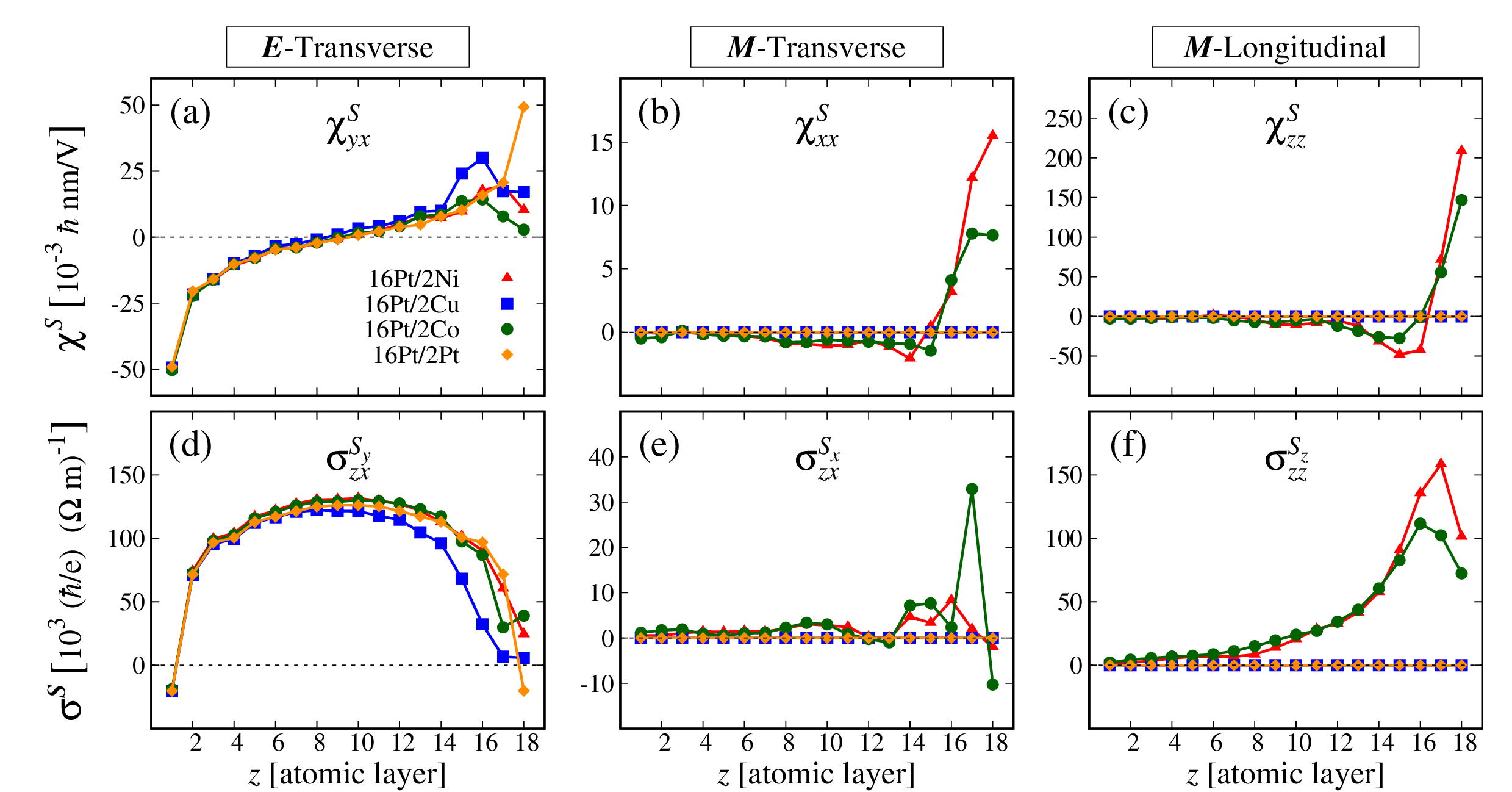}
\caption{Computed atomic layer-resolved nonzero components of the spin ME susceptibility $\boldsymbol{\chi}^{{S}}$ and spin conductivity $\boldsymbol{\sigma}^{\bm{S}}$ of the 16Pt/2$Y$ structures.
(a) The $\bm{E}$-transverse component $\chi_{yx}^{S}$, (b) $\bm{M}$-transverse component $\chi_{xx}^{S}$, and (c) $\bm{M}$-longitudinal  component $\chi_{zz}^{S}$.
The corresponding components of the spin conductivity tensor are given as (d) $\bm{E}$-transverse $\sigma^{S_y}_{zx}$, (e) $\bm{M}$-transverse $\sigma^{S_x}_{zx}$, and (f) $\bm{M}$-longitudinal $\sigma^{S_z}_{zz}$. 
The $\bm{E}$-transverse components {\blue of $\bm{\sigma}^S$} can be associated with SHE.
The $\bm{M}$-transverse
components {\blue of $\bm{\chi}$} are nonzero only for magnetic systems (16Pt/2Ni and 16Pt/2Co) and in the vicinity of the interface, suggesting 
{\blue the importance of} spin-splitting of the electronic states. %and can thus be associated with the SREE. 
The $\bm{M}$-longitudinal components are discussed in the text. See Fig.~\ref{fig:16Pt2Y_Schematic} for the numbering of the atomic layers.}
\label{fig:16Pt2Y_SS_SC}
\end{figure*}

\begin{figure*}[t!]
\centering
\includegraphics[width=0.85\linewidth]{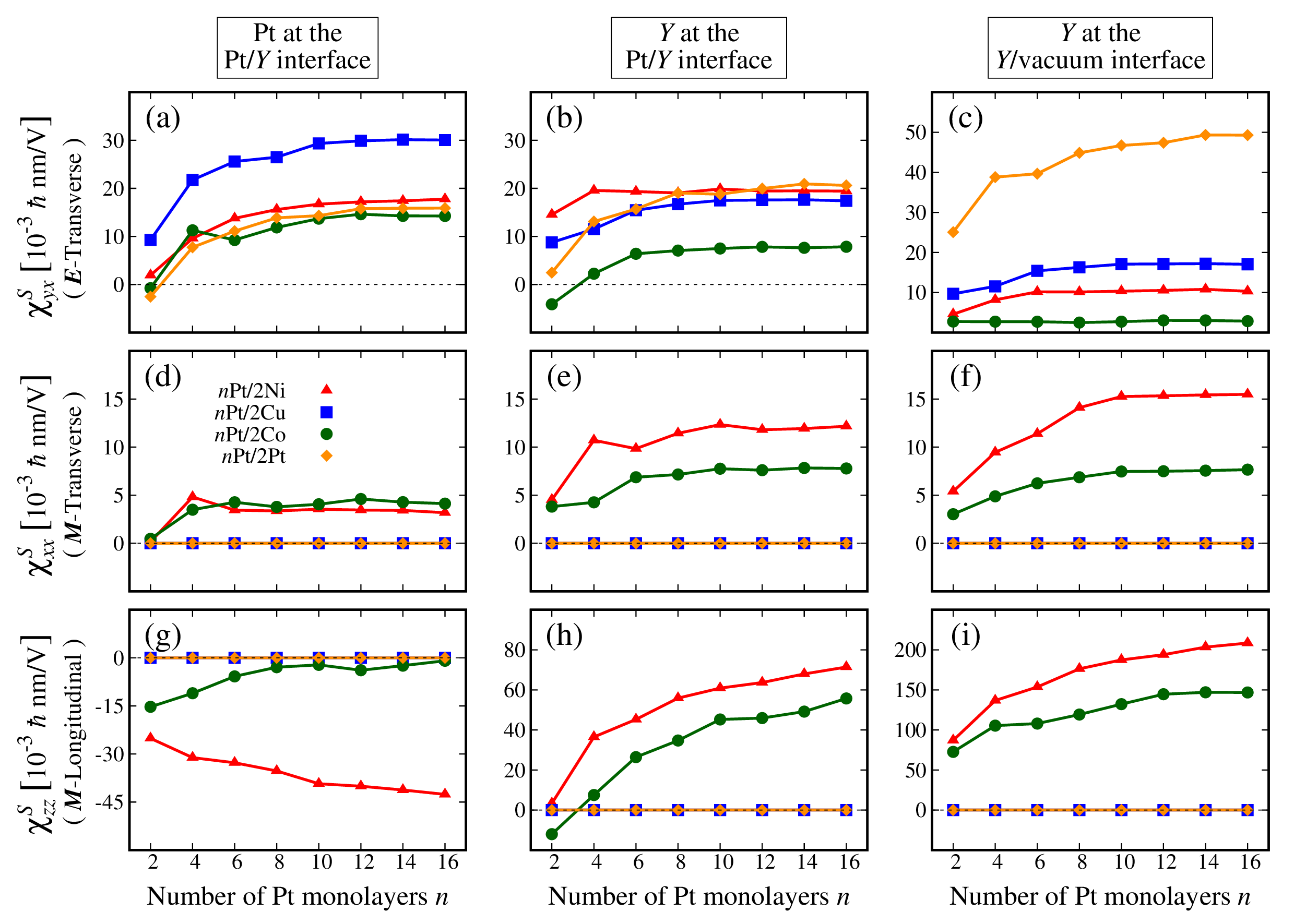}
\caption{Dependence of the spin ME susceptibility tensor $\boldsymbol{\chi}^S$ on the number of Pt monolayers $n$. Top row: $\bm{E}$-transverse component of $\boldsymbol{\chi}^S$, for (a) the Pt atom at the Pt/$Y$ interface, (b) the $Y$ atom at the Pt/$Y$ interface, and (c) the $Y$ atom at the $Y$/vacuum interface. Second row: $\bm{M}$-transverse component of $\boldsymbol{\chi}^S$, for (d) the Pt atom at the Pt/$Y$ interface, (e) the $Y$ atom at the Pt/$Y$ interface, and (f) the $Y$ atom at the $Y$/vacuum interface. Bottom row: $\bm{M}$-longitudinal component of $\boldsymbol{\chi}^S$, for (g) the Pt atom at the Pt/$Y$ interface, (h) the $Y$ atom at the Pt/$Y$ interface, and (i) the $Y$ atom at the $Y$/vacuum interface.}
\label{fig:NPt2Y_thickness_study}
\end{figure*}

%%%%%%%%%%%%%%%%%%%%%%
%%% RESULT SECTION %%%
%%%%%%%%%%%%%%%%%%%%%%

\section{Results}
\subsection{Spin response}
\subsubsection{Magnetization out-of-plane}
We start with the case where $\bm{M} \, || \, \bm{u}_z$. It is instructive to consider first the thickest heterostructures, i.e., 16Pt/2Ni, 16Pt/2Co, 16Pt/2Cu, and 16Pt/2Pt.
In Figs.\ \ref{fig:16Pt2Y_SS_SC}(a), (b), and (c) we show the computed atomic layer-resolved profiles of the aforementioned nonzero components of $\boldsymbol{\chi}^S${\blue, as well as related components of the spin conductivity tensor $\sigma^{S_y}_{zx}$, $\sigma^{S_x}_{zx}$, and $\sigma^{S_z}_{zz}$ in Figs.~\ref{fig:16Pt2Y_SS_SC}(d), (e), and (f), respectively.} 

In all cases, we observe that the response of the Pt atomic-layer at the vacuum interface  ($z=1$) is virtually independent on the type of $Y$ atom used, suggesting that these systems are thick enough to isolate the Pt/$3d$-interface properties. The inclusion of the two monoatomic layers of $3d$ elements mainly impacts the $\boldsymbol{\chi}^S$ and $\boldsymbol{\sigma}^{\bm{S}}$ profiles close to their interface.

For the $\bm{E}$-transverse components (Figs.\ \ref{fig:16Pt2Y_SS_SC}(a) and (d)), both $\boldsymbol{\chi}^S$ and $\boldsymbol{\sigma}^{\bm{S}}$ are qualitatively barely impacted {\blue in the bulk Pt region} by the replacement of the two last Pt atomic monolayers by two $3d$ atomic monolayers. The profile of $\sigma^{S_y}_{zx}$ is in all cases mostly defined by a plateau in the center of the Pt layer. {\blue This specific component of $\boldsymbol{\sigma}^{S}$ is customarily identified with the SHE in Pt-bulk calculations.} {\blue The spin-accumulation profile related to $\chi^S_{yx}$ across the Pt layer strongly resembles the type spin accumulation that is expected from transverse spin flow. 
{\blue Reversing the magnetization of the Ni and Co layers from $+\bm{u}_z$ to $-\bm{u}_z$ in the calculations, moreover, does not have a notable effect on the ($\bm{M}$-even) spin-accumulation given by $\chi_{yx}^S$.}
These results are in agreement with what is observed in SHE calculations for bulk Pt \cite{Zhang2000,Stamm2017}. Note that the accumulated spin moment is given by $\delta \bm{M}= -2 ({\mu_B}/\hbar) \delta \bm{S}$.}

{\blue For the 16Pt/2Cu system, the discussion cannot be performed on the basis of torques ($\bm{M}$ is ill-defined). For the spin accumulation related to $\chi^S_{yx}$, compared to the pure Pt case, we observe an increase of accumulated spin density at the Pt side of the Pt/Cu interface as well as a decrease of spin density within the Cu layer.}

For 16Pt/2Ni and 16Pt/2Co one can in addition observe that a spin depolarization 
occurs in the two ferromagnetic layers, as has been discussed in Refs.~\cite{Kurt2003,Rojas-Sanchez2014,Nguyen2014,Dolui2017,Tao2018}.

 The $\bm{M_{\bot}}$ components (Figs.\ \ref{fig:16Pt2Y_SS_SC}(b) and (e)), are nonexistent for non-magnetic pure Pt and Pt/Cu system. Sizable values are only obtained close to the interface with Co and Ni, both for $\boldsymbol{\chi}^S$ and $\boldsymbol{\sigma}^{\bm{S}}$. Remarkably, while the $\bm{E}_\bot$ and $\bm{M}_\bot$ components are comparable in size close to the interface, their features differ greatly: (1) there is no bulk-like behavior for $\sigma^{S_y}_{zy}$/$\sigma^{S_x}_{zx}$, (2) those components are non-existent in bulk Pt, and (3) they are magnetization direction dependent. 
Although there is a spatial symmetry breaking at the Pt/Cu interface, no spin polarization is induced, {\blue which strongly suggests the importance of spin-split electronic states at the interface for $\chi_{xx}^S$ and $\sigma_{zx}^{S_x}$.}
{\blue The odd-in-$\bm{M}$ spin conductivity $\sigma_{zx}^{S_x}$ was recently named magnetic SHE \cite{Kimata2019}, to distinguish it from the conventional SHE $\sigma_{zx}^{S_y}$ that is even-in-$\bm{M}$ \cite{Chuang2020}.

{\blue As aforementioned}, the $\bm{M}$-longitudinal components (see Figs.\ \ref{fig:16Pt2Y_SS_SC}(c) and (f)) are peculiar in the sense that they are not SOT-related (the usual SOT configuration does not involve out-of-plane electrical fields and, also, no torque is generated by a spin accumulation parallel to the static moment). Although such components can {\blue in principle} be obtained via a symmetry analysis (cf.\ Ref.~\cite{Zelezny2017}), they haven't been, to the best of our knowledge, investigated so far. %Nonetheless, 
As will be clarified further below, this effect is due to the spin-orbit interaction. Here, an electric field applied parallel to the out-of-plane magnetization causes a sign-changing spin polarization along $\bm{M}$ in the $\sim$5 topmost monolayers. This is clearly a magnetic effect, as it does not exist for the nonmagnetic systems. The spin conductivity $\sigma_{zz}^{S_z}$ shows a decaying behavior from $z=16$ to $z=1$, but this decay is slower than that of the equilibrium spin magnetization in the systems. Also, similar to the $\bm{M}$-transverse component, no ``bulk-like" behavior is observed. A possible way of observing this previously unidentified SOC-induced effect could be achieved by gating the ferromagnetic layer from the top and monitor a change of its magnetization.

So far, we focused on the components of the spin conductivity tensor giving rise to spin currents flowing along $\bm{u}_z$. While those components are the ones that should be of interest for understanding SOT in bilayer structures, other nonzero components can be observed, as well. For a magnetic system with the $4mm$ point group, {\violet with $\bm{\mathcal{M}} \parallel \bm{u}_{z}$}, we find that $\boldsymbol{\sigma}^{\bm{S}}$ can be written as 
\begin{equation}
\begin{split}
\boldsymbol{\sigma}^{S_x} &= 
\begin{pmatrix}
0 & 0 & \sigma^{S_x}_{xz} \\[0.3em]
0 &  0 & \sigma^{S_x}_{yz} \\[0.3em]
\sigma^{S_x}_{zx} & \sigma^{S_x}_{zy}& 0
\end{pmatrix}, \nonumber
\end{split}
\end{equation}
\begin{equation}
\begin{split}
\boldsymbol{\sigma}^{S_y} &= 
\begin{pmatrix}
0 & 0 & \sigma^{S_y}_{xz} \\[0.3em]
0 &  0 & \sigma^{S_y}_{yz} \\[0.3em]
\sigma^{S_y}_{zx} & \sigma^{S_y}_{zy}& 0
\end{pmatrix},
\end{split}
\end{equation}
\begin{equation}
\begin{split}
\boldsymbol{\sigma}^{S_z} &= 
\begin{pmatrix}
\sigma^{S_z}_{xx} & \sigma^{S_z}_{xy} & 0 \\[0.3em]
\sigma^{S_z}_{yx} & \sigma^{S_z}_{yy} &  0 \\[0.3em]
0 & 0 & \sigma^{S_z}_{zz}
\end{pmatrix}.
\nonumber
\end{split}
\end{equation}

The components associated to the SHE, i.e., $\sigma^{S_k}_{ij}$, where the indices are such that $\epsilon_{ijk} \neq 0$ ($\epsilon_{ijk}$ is the Levi-Civita symbol), are nonzero {\violet whether the system is magnetic or not.} However, while in cubic systems like bulk Pt they are all equal in magnitude, here, because of the symmetry breaking with respect to the $z$ axis, the tensor elements are not invariant under exchange of $z$ and $x$ or $y$ indices.
{\violet The other 7 components, that is $\sigma^{S_x}_{xz}$, $\sigma^{S_x}_{zx}$, $\sigma^{S_y}_{yz}$, $\sigma^{S_y}_{zy}$, $\sigma^{S_z}_{xx}$, $\sigma^{S_z}_{yy}$, and $\sigma^{S_z}_{zz}$ only exists for magnetic systems. Some of those magnetic components have  been discussed recently \cite{Humphries2017,Chuang2020,TaoWang2020,Qu2020,Safranski2020}.}

\subsubsection{Pt-thickness dependence}
Next, we investigate the Pt layer thickness dependence of the  $\bm{E}_{\bot}$, $\bm{M}_{\bot}$, and $\bm{M}_{\parallel}$ components of $\boldsymbol{\chi}^S$. The number of Pt monolayers for our $n$Pt/$2Y$ systems is varied from $n=2$ (Pt thickness $\sim 0.38 $ nm) to $n=16$ (Pt thickness $\sim 3.08 $ nm). Figure \ref{fig:NPt2Y_thickness_study} shows the computed Pt-thickness dependence where each column of the figure focuses on one particular atomic monolayer, with, from left to right, the Pt monolayer at the Pt/$Y$ interface, the $Y$ monolayer at the Pt/$Y$ interface, and the $Y$ atomic monolayer at the $Y$/vacuum interface. Each row focuses on one particular component, namely, from top to bottom, the $\bm{E}_{\bot}$, $\bm{M}_{\bot}$, and $\bm{M}_{\parallel}$ components. The values of the tensor elements that give rise to the SOT, the $\bm{E}_{\bot}$ and $\bm{M}_{\bot}$ components, barely fluctuate beyond $n=8$ (Pt thickness $\geq 1.54 $ nm). {\blue Thus, both $\bm{\mathcal{T}}_\text{FL}$ and $\bm{\mathcal{T}}_\text{DL}$ approach their maximum values already for relatively thin Pt layers,} {\blue consistent with recent investigations \cite{Nguyen2016,Mahfouzi2020}.}

For the $\bm{E}_{\bot}$ components, in the case of pure Pt ($n$Pt/$2$Pt), $\chi_{yx}^S$ tends to increase the closer we come to the last layer, which is typically what we would expect from a  {\blue transport}-generated spin accumulation profile. When the two last layers are replaced by a magnetic element ($Y=$ Co or Ni) drastic changes occur. First, we observe that $\chi_{yx}^S$ is bigger for the $3d$ monolayer closer to the Pt layer than for the second $Y$ layer. 
Second, at a fixed position, $\chi_{yx}^S$ is bigger for $Y=$ Ni than $Y=$ Co.

For the $\bm{M}_{\bot}$ components, {\blue associated to $\bm{\mathcal{T}}_\text{DL}$,} the $\chi_{xx}^S$ for the Pt monolayer at the Pt/$Y$ interface (Fig.\ \ref{fig:NPt2Y_thickness_study}(d)) is virtually identical for $Y=$ Ni or Co and for all Pt thicknesses considered. In the first and second magnetic monolayer  (Figs.\ \ref{fig:NPt2Y_thickness_study}(e)) and (f))
the $\chi_{xx}^S$ is, in both cases, bigger for $Y=$ Ni than $Y=$ Co. The bottom row, lastly, shows the $\bm{M}$-longitudinal spin accumulation. Also here, we obtain that a larger magnitude of $\chi_{zz}^S$ is generated for $Y$ = Ni.

\subsubsection{Magnetization-direction dependence of the spin response}
The direction of the magnetization vector $\bm{M}$ is often rotated in out-of-plane or in-plane directions in experiments
\cite{Avci2014,Du2020}. {\blue The magnetization-direction unit vector $\bm{\mathcal{M}}$  can be written as
$ \bm{\mathcal{M}} = (\sin\theta \cos\phi, \sin\theta  \sin\phi, \cos\theta)$,
where $\phi$ and $\theta$ are the azimuthal and polar angles, respectively. To capture the $\bm{\mathcal{M}}$-dependence of $\bm{\chi}^S$, we self-consistently compute $\bm{\chi}^S$ at different ($\theta$, $\phi$) values.} We choose the 12Pt/2Co system for these simulations and consider the ME spin response on the first Co atom at the interface with the Pt layer. {\blue In particular, we compute $\bm{\chi}^S(\theta, \phi)$ for $\bm{\mathcal{M}}$ constrained in the $x-y$ plane (Fig.~\ref{fig:Chi_S_inplane}) and $\bm{\mathcal{M}}$ constrained in the $x-z$ plane (Fig.~\ref{fig:Chi_S_outplane}).}

\begin{figure}[thp!]
\centering
\includegraphics[width=0.99\linewidth]{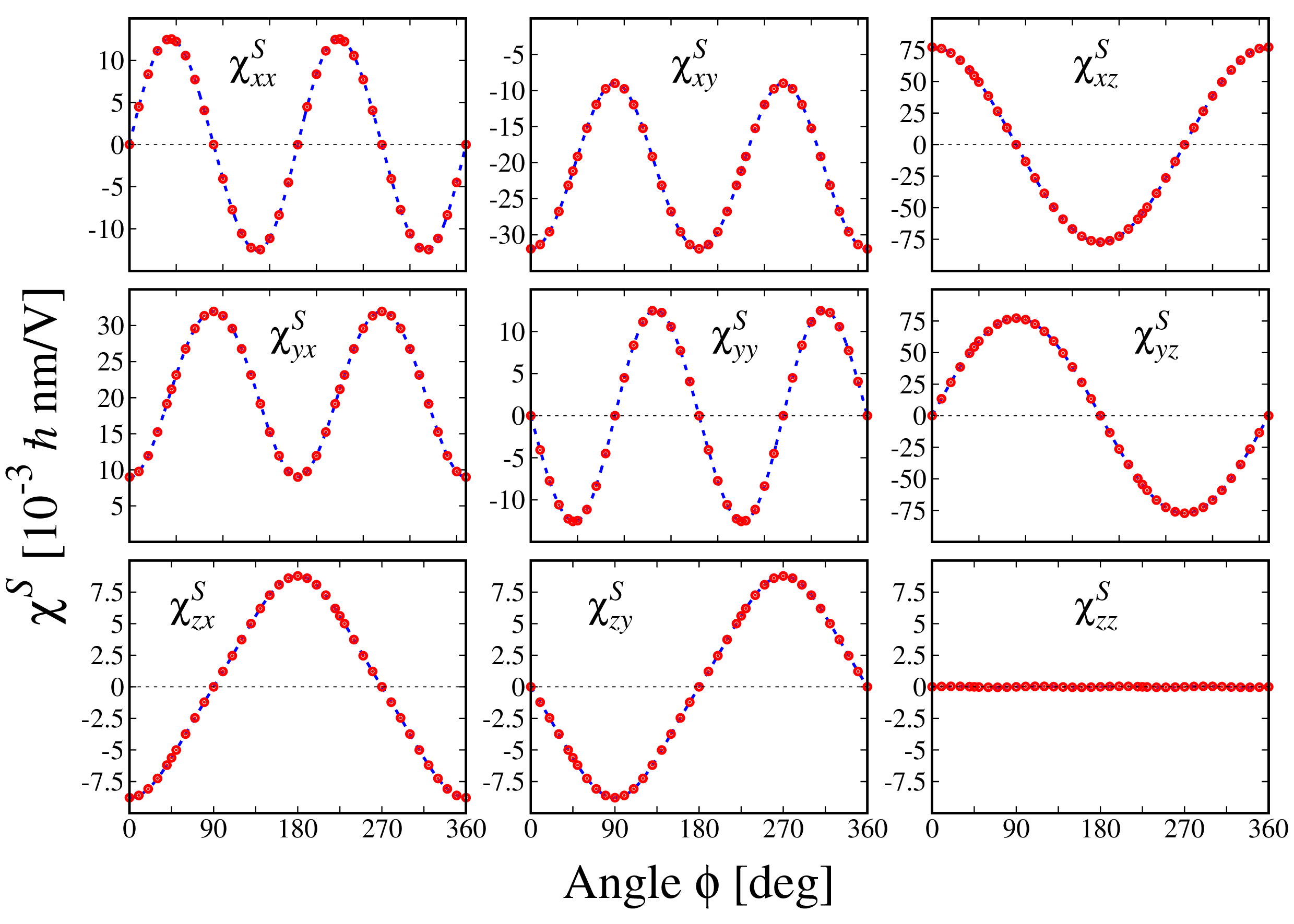}
\caption{Computed angular dependence of the spin ME tensor $\bm{\chi}^S$ of the 12Pt/2Co system {\blue at the first Co atom} for in-plane magnetizations, given as a function of the azimuthal angle $\phi$.}
\label{fig:Chi_S_inplane}
\end{figure}

In Fig.\ \ref{fig:Chi_S_inplane} we show the computed angular dependence of the tensor components when the magnetization is in-plane and rotated from angle $\phi =0 ^{\circ}$ (i.e., parallel to $\bm{u}_x$) to $360^{\circ}$. The panels reveal that spin ME tensor is, to great precision, given by the following parametric form,
\begin{equation}
\boldsymbol{\chi}^S = 
\begin{pmatrix}
\alpha_1 \sin 2\phi & \alpha_2 + \alpha_3\cos 2\phi  & \alpha_4 \cos \phi \\[0.3em]
-\alpha_2 -\alpha_3 \cos 2\phi  & -\alpha_1\sin 2\phi  &  \alpha_4 \sin \phi \\[0.3em]
\alpha_5 \cos \phi & \alpha_5 \sin \phi  & 0
\end{pmatrix} ,
\label{chi-x-phi}
\end{equation}
where $\alpha_i$ ($i=1,\, 2,\cdots ,\,5$) are constants in units of $10^{-3} \hbar \frac{\textrm{nm}}{\textrm{V}}$, whose values are
$ \alpha_1 = 12.54$,  $\alpha_2 = -20.83$, 
$\alpha_3 = 11.50$,
$\alpha_4 = 77.18$, and
$\alpha_5 = -8.36$.
Note that this functional form only gives the $\phi$-dependence, {\blue that is, $\theta$ is kept constant to $90^{\circ}$}. 
The computed angle dependence illustrates that the tensor elements, and thus the resulting SOTs, obey a distinct angular symmetry. {\blue Terms whose angular dependency varies as $2\phi$ are even under magnetization reversal, while those that vary as $\phi$ are odd.}

Let us consider for example the case where $\bm{M}$ is along $\bm{u}_x$, that is $\phi=0$. If we apply a field $\bm{E}$ along $\bm{M}$, the only relevant nonzero elements, {\blue that is, the elements that give rise to a torque}, are $\chi_{yx}^S$ and $\chi_{zx}^S$. The former tensor element is even in the magnetization, giving consistently a `$-\alpha_2-\alpha_3 \cos 2\phi$' angular dependence, while the latter element is odd in $\bm{M}$, giving a $\cos \phi $ dependence. Note that the expression (\ref{eq:SOT-T-expression}) for the SOT contains the cross product $\bm{M} \times (\bm{\chi}^S \bm{E})$, which implies that an additional trigonometric function ($\sin \phi$, $\cos \phi$) appears in the SOT.

\begin{figure}[thp!]
\centering
\includegraphics[width=0.99\linewidth]{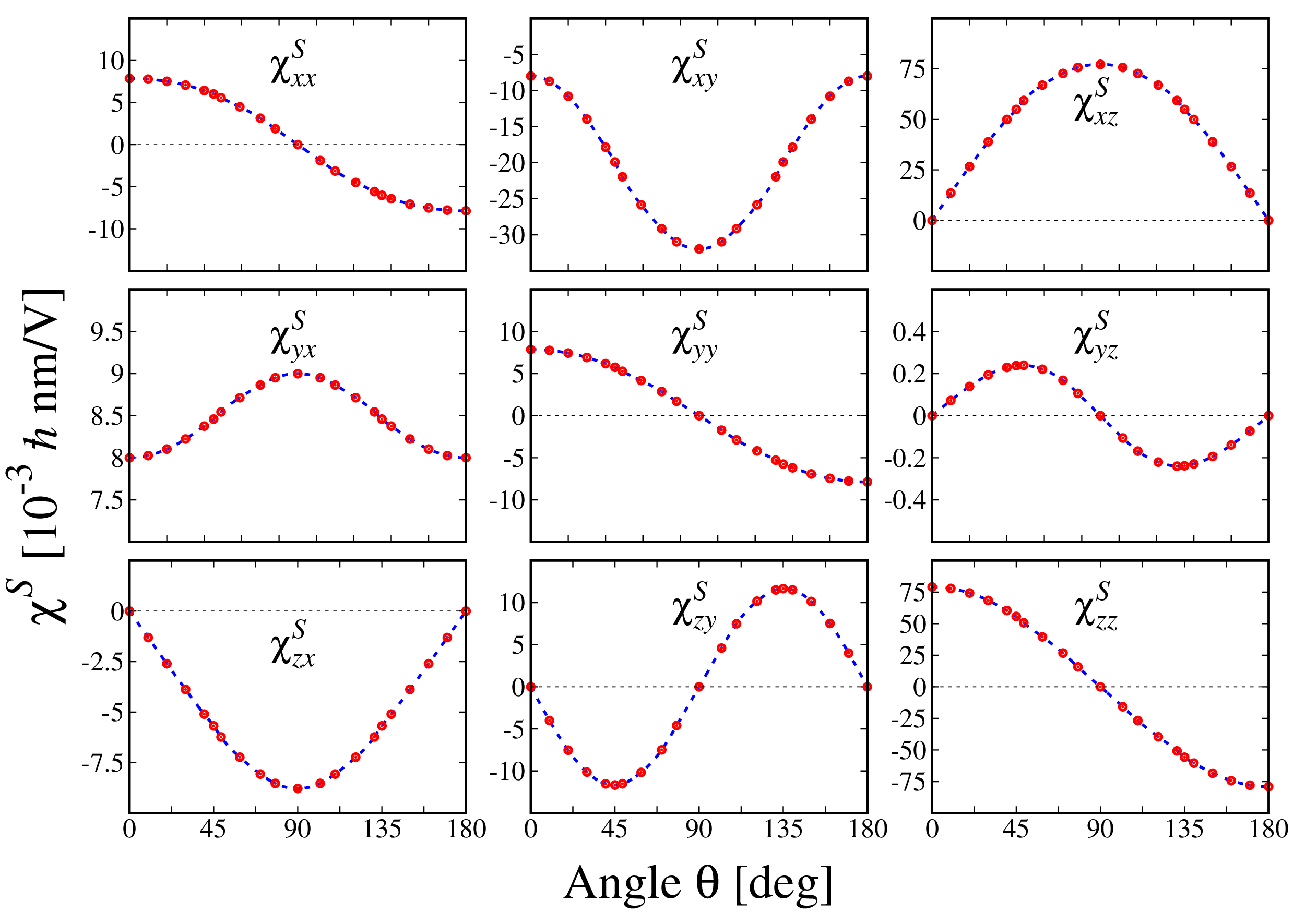}
\caption{Computed angular dependence of the spin ME tensor $\bm{\chi}^S$ of the 12Pt/2Co system {\blue at the first Co atom} for magnetizations in the $x - z$ plane, given as a function of the polar angle $\theta$ (and for $\phi=0$).}
\label{fig:Chi_S_outplane}
\end{figure}

Analogous calculations have been carried out for $\bm{M}$ being varied along the polar angle $\theta$, {\blue with $\phi$ set to 0}. Figure \ref{fig:Chi_S_outplane} shows the computed $\theta$-angle dependence of the ME tensor $\bm{\chi}^S$ for the 12Pt/2Co system. It can be recognized that the angular variation of the matrix elements obeys the $\theta$-dependence
\begin{equation}
\boldsymbol{\chi}^S \propto
\begin{pmatrix}
\cos \theta & \beta_1 + \cos 2\theta  &  \cos \theta \\[0.3em]
\beta_2 - \cos 2\theta  &  \cos \theta &   \sin 2\theta \\[0.3em]
 - \sin \theta &  - \sin 2\theta  &  \cos \theta
\end{pmatrix} ,
\label{chi-z-theta}
\end{equation}
where $\beta_i$ and $\beta_2$ are constants. Here, we only give the generic trigonometric dependence of the tensor elements. In this case there are nine constants needed to describe the angle dependence precisely as in Eq.\ (\ref{chi-x-phi}).

\begin{figure}[b!]
\centering
\includegraphics[width=0.8\linewidth]{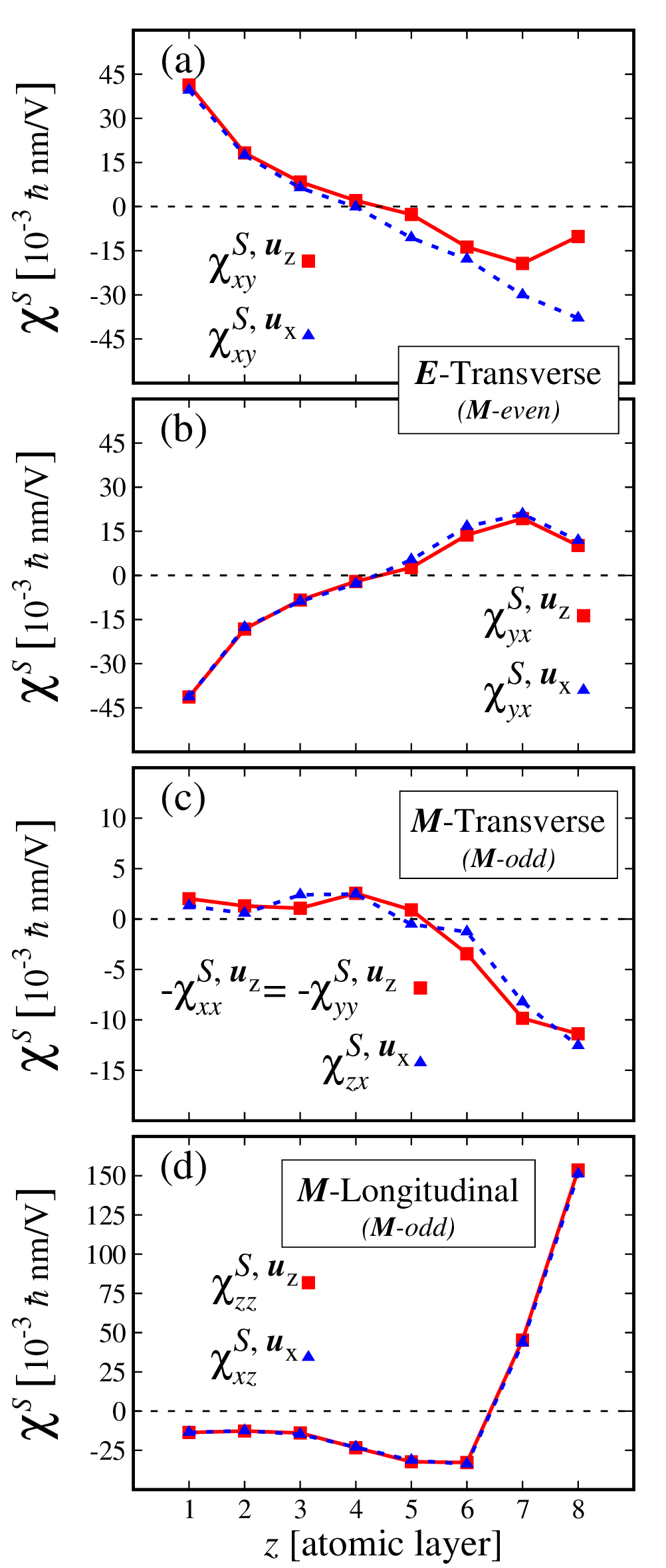}
\caption{Magnetization-direction dependence of the spin ME susceptibility tensor $\boldsymbol{\chi}^{S}$ for $6$Pt/$2$Ni. Calculated data are given for $\bm{M} \,|| \, \bm{u}_z$ ($\bm{M} \, || \, \bm{u}_x$) by the red squares (blue triangles). When the magnetization direction switches from $\bm{u}_z$ to $\bm{u}_x$, the transverse components $\chi_{xy}^{S,\bm{u}_z}$ and $\chi_{yx}^{S,\bm{u}_x}$  are mapped onto themselves, with $\chi_{xy}^S$ being notably modified close to the Pt/Ni interface while $\chi_{yx}^S$ is barely affected. The $\bm{M}$-transverse components $\chi_{xx/yy}^{S,\bm{u}_z}$ 
are mapped onto $-\chi_{zx}^{S,\bm{u}_x}$.
The $\bm{M}$-longitudinal $\chi_{zz}^{s,\bm{u}_z}$ component
is transformed onto $\chi_{xz}^{S,\bm{u}_x}$.}
\label{fig:6Pt2Ni_SS_Mz_Mx_Vertical}
\end{figure}

{\blue The angular dependence of the $\chi^S$ tensor
was computed for a particular atomic position 
in the bilayer. Performing further calculations, we find that the obtained {\blue angular} dependence [Eqs.\ (\ref{chi-x-phi}) and (\ref{chi-z-theta})] is valid at any atomic position in the layer,  albeit with vanishing constants for the odd-in-$\bm{M}$ components when the atomic layer is close to the vacuum/Pt interface. For each atomic position there are different constants for the tensor components, however, a decisive property is, whether the tensor component is $\bm{E}$-transverse, $\bm{M}$-transverse, or $\bm{M}$-longitudinal. As illustrated in Table \ref{tab:MSymmetry}, in this way equivalent tensor elements can be identified for different $\bm{M}$ directions.}

In Fig.\ \ref{fig:6Pt2Ni_SS_Mz_Mx_Vertical}  we compare the atomic-layer resolved profiles of $\boldsymbol{\chi}^S$ for the 6Pt/2Ni system, for  $\bm{M} \, || \, \bm{u}_x$ and $\bm{M} \, || \, \bm{u}_x$. We compare the $\bm{E}_{\bot}$, $\bm{M}_{\bot}$ and $\bm{M}_{\parallel}$ components, which allows us to track the physically equivalent quantities with respect to the SOT symmetry (FL or DL).
We use in Fig.\ \ref{fig:6Pt2Ni_SS_Mz_Mx_Vertical} the superscript $\bm{u}_z$ ($\bm{u}_x$) to denote quantities computed with $\bm{M} \, || \, \bm{u}_z$ ($\bm{M} \, || \, \bm{u}_x$).
{\blue 
For most of the atomic positions, the corresponding tensor elements for both $\bm{M} \, || \, \bm{u}_z$ and $\bm{M} \,||\, \bm{u}_x$ show a strikingly similar behavior, with the exception of the $\chi^S_{xy}$ element that differs most in the Ni layer. The similar behavior and approximately identical values illustrate that the tensor elements transform in an almost rigid manner when the magnetization vector is rotated. The larger difference in the 
 $\chi_{xy}^S$ component suggest that the orientation of $\bm{M}$ has a stronger influence on the Pt/Ni interface electronic structure. 
 {\blue Note that although $\chi_{xy}^{S,\bm{u}_x}$ and $\chi_{xy}^{S,\bm{u}_z}$ differ significantly at the Pt/Ni interface (see Fig.~\ref{fig:6Pt2Ni_SS_Mz_Mx_Vertical}(a)), $\chi_{xy}^{S,\bm{u}_x}$ does not become effective, as it would give rise to an induced magnetization along the equilibrium magnetization, and can therefore not contribute to SOT.}

For the $\bm{M}_{\bot}$ components, shown in Fig.~\ref{fig:6Pt2Ni_SS_Mz_Mx_Vertical}(c), very similar magnitudes are observed but there is an opposite sign in the $\chi_{xx}^{S,\bm{u}_z}$ and $\chi_{zx}^{S,\bm{u}_x}$ components.
 This sign reversal is easily recovered using our proposed classification. Indeed, in the case $\bm{M} \, || \, \bm{u}_z$, one finds from Eq.\ (\ref{eq:Symmetry_Relation_Mtrans}) for the corresponding component
$$
\chi_{xx}^{\bm{u}_z} \hspace{0.1cm} : \hspace{0.1cm}
\delta \bm{S} \, \propto \, (\bm{u}_x \times \bm{u}_z) \times \bm{u}_z = - \bm{u}_x \,,
$$
while for $\bm{M} \, || \, \bm{u}_x$ we have,
$$
\chi_{zx}^{\bm{u}_x} \hspace{0.1cm} : \hspace{0.1cm}
\delta \bm{S} \, \propto \, (\bm{u}_x \times \bm{u}_z) \times \bm{u}_x = + \bm{u}_z \,,
$$
which perfectly captures the sign reversal.}
The $\bm{M}$-longitudinal $\bm{\chi}^S$ components, lastly, follow the approximate transformation behavior quite well, see Fig.\ \ref{fig:6Pt2Ni_SS_Mz_Mx_Vertical}(d). 
These calculations illustrate the usefulness of the classification of the tensor elements according $\bm{E}_{\bot}$, $\bm{M}_{\bot}$, and $\bm{M}_{||}$ for identifying their resulting torque properties.

\subsubsection{Electronic lifetime dependence}
\label{Sec:lifetime-dep}

\begin{figure}[tbh!]
\centering
\includegraphics[width=0.70\linewidth]{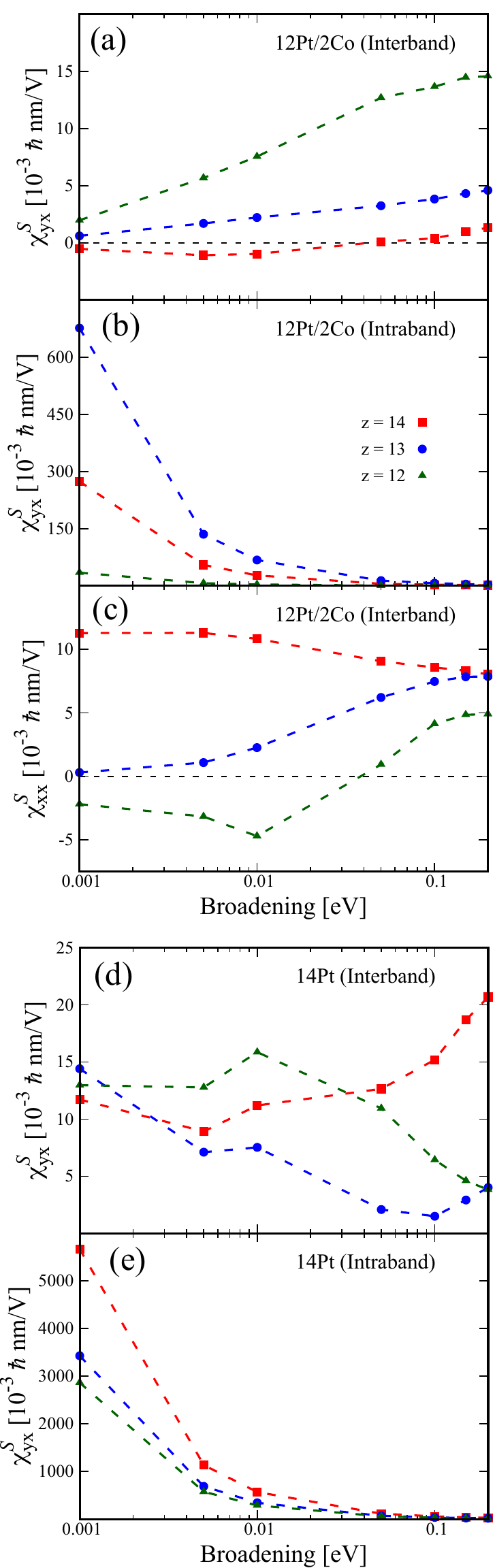}
\caption{Dependence of the interband and intraband contributions of the spin ME response $\chi^S$ on the electronic broadening $\delta$, computed for the 12Pt/2Co and 14Pt systems. The $\chi_{xx}^S$ component is purely interband while $\chi_{xy}^S$ contains both interband and intraband contributions.
The electronic broadening used in the main part of this article, $\delta = 0.22 - 0.27$ eV, is a realistic value for metallic systems.}
\label{fig:broadening_dep_vertical_2}
\end{figure}

{\blue In the above calculations we employed the intraband and interband electron lifetimes ($\hbar \tau_{intra} = 0.220$ eV, $\hbar \tau_{inter} = 0.272$ eV) obtained from fitting the calculated linear-response conductivity to experimental electrical conductivities of Pt films \cite{Stamm2017}. In general, these electronic lifetimes depend on the sample purity and microstructure. To investigate the influence of these lifetimes on the nonequilibrium spin and orbital polarizations we have varied these parameters in our calculations, adopting the 12Pt/2Co and 14Pt films as typical model systems. We consider the case $\bm{M} \, || \, \bm{u}_z$ for the 12Pt/2Co film, for which the nonzero spin tensor elements are $ \chi_{yx}^S$, $\chi_{xx}^S$, and $\chi_{zz}^S$. For the 14Pt film only the tensor element $\chi_{yx}^S$ is nonzero, cf.\ the results shown in Fig.\ \ref{fig:16Pt2Y_SS_SC}.
At this point it is relevant to mention that the $\bm{M}$-transverse tensor element $\chi_{xx}^S$ has only \textit{interband} contributions, whereas the $\bm{E}$-transverse spin polarization $\chi_{yx}^S$ has both \textit{intra}- and \textit{inter}-band contributions. 
The calculated dependence of the torque-related, spin ME responses on the electronic broadening $\hbar \delta$ ($=\hbar/\tau$) is shown in Fig.\ \ref{fig:broadening_dep_vertical_2}. The intraband part of $\chi_{yx}^S$ increases linearly for decreasing lifetime broadening, as expected from Eq.\ (\ref{eq:LinResp}). The interband contributions, conversely, vary with the lifetime broadening but become more broadening independent for large values of $\delta$. For small $\delta$ the interband contributions converge to nonzero values (but the values for $\chi_{yx}^S$ become small). For realistic lifetime values ($\hbar/\tau \sim 0.25$ eV) the intra- and interband induced spin polarizations have the same magnitude and need to be treated on equal footing.  Note that semiclassical Boltzmann transport theory only captures the intraband spin ME component \cite{Johansson2018}, 
an approximation that is viable for very pure crystals with long electron lifetimes. 

In the limit of very small electronic broadenings, 
the intraband contribution to $\chi_{yx}^S$ will dominate completely the electrically induced spin polarization. This is the component that leads to a fieldlike SOT, see Table \ref{tab:MSymmetry}. The intraband part of the spin response is commonly associated with the Rashba-Edelstein effect \cite{Zelezny2017,Kimata2019}, consistent with the shape of the Rashba spin-orbit coupling Hamiltonian \cite{Bychkov1984} that leads to a fieldlike torque in Landau-Lifshitz-Gilbert equations of spin dynamics. 
The spin Hall effect is conversely commonly associated with an interband contribution \cite{Sinova2015}. Such contribution is present both in $\chi_{yx}^S$ and $\chi_{xx}^S$, i.e., in the $E_{\bot}$ and $M_{\bot}$ terms. The SHE will thus contribute to both a fieldlike and a dampinglike SOT. The SHE is considered to lead to a DL torque \cite{Liu2011} when using the Slonczewski model \cite{Slonczewski1996}. The SHE was however also proposed to lead to a FL torque \cite{Haney2013,Amin2016}, consistent with our calculations.
{\blue The latter identification was made within the microelectronic circuit model, wherein the $\bm{M}_{\bot}$ ($\bm{E}_{\bot}$) component is mainly due to the SHE (SREE), when the spin mixing conductance is chiefly real \cite{Haney2013,Amin2016}. To obtain such result, it is assumed that the transverse spin current and spin accumulation 
exists in the HM layer only, i.e., these quantities are
 zero in the FM layer. This might be a reasonable approximation for thicker layers, but in our case, where we compute atomistic quantities, we find a nonzero spin conductivity and spin accumulation (see Fig.\ \ref{fig:16Pt2Y_SS_SC}) in the FM layer which is only two atoms thick. {\blue For a very small lifetime broadening the FL $\bm{E}_{\bot}$ contribution will dominantly stem from the intraband SREE and the DL $\bm{M}_{\bot}$ contribution from the SHE.}}
We note however with respect to this discussion, that our DFT Hamiltonian contains the full form of the spin-orbit interaction and is thus different from the more elementary Bychkov-Rashba SOC \cite{Bychkov1984}, but it provides all materials' specific SOC effects.

}

\begin{figure*}[ht!]
\centering
\includegraphics[width=\linewidth]{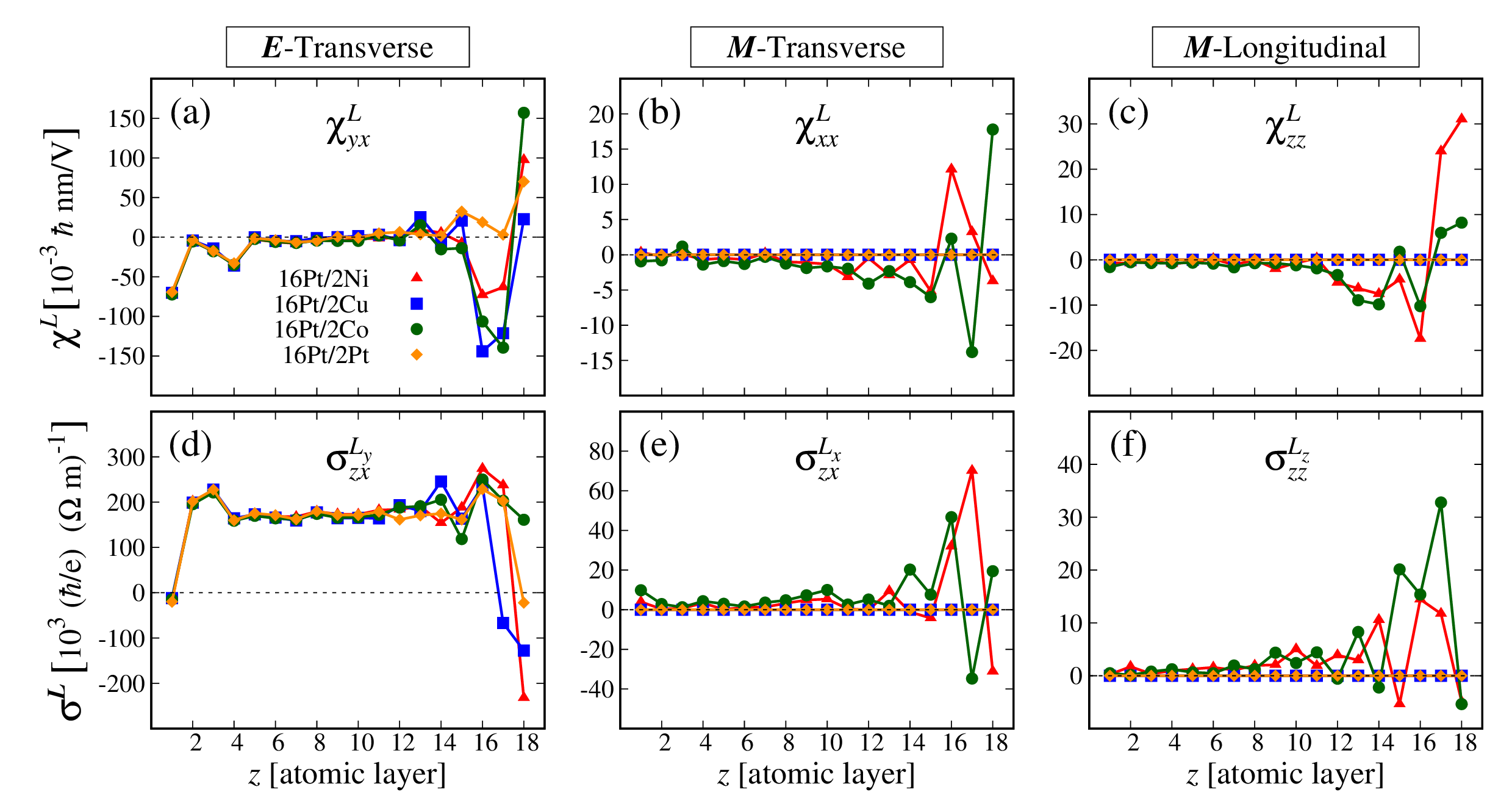}
\caption{Atomic layer-resolved nonzero components of the orbital ME susceptibility $\bm{\chi}^L$ and the orbital conductivity $\bm{\sigma}^{\bm{L}}$ of the 16Pt/2$Y$ films with $\bm{M} \, || \, \bm{u}_z$. (a) The $\bm{E}$-transverse component $\chi_{yx}^L$, (b) $\bm{M}$-transverse component $\chi_{xx}^L$, and (c) the $\bm{M}$-longitudinal component $\chi_{zz}^L$.
The corresponding components of the orbital conductivity tensor are given as (d) $\bm{E}$-transverse $\sigma^{L_y}_{zx}$, (e) $\bm{M}$-transverse  $\sigma^{L_x}_{zx}$, and (f) $\bm{M}$-longitudinal $\sigma^{L_z}_{zz}$.
The $\bm{E}$-transverse component $\sigma_{zx}^{L_y}$ is associated with the OHE conductivity, conventionally defined for bulk Pt. The $\bm{M}$-transverse components are nonzero only for the magnetic systems (16Pt/2Ni and 16Pt/2Co),  and also
%and are associated with the OREE. 
the $\bm{M}$-longitudinal components arise from the spin polarization of the electronic states.}
\label{fig:16Pt2Y_OS_OC}
\end{figure*}

\subsection{Orbital response}

\subsubsection{Atomic layer-resolved orbital response}

A similar analysis can be performed for the orbital response, both in terms of $\boldsymbol{\chi}^{L}$ and $\boldsymbol{\sigma}^{\bm{L}}$. While similarities are observed, unique characteristic can be observed, too.
{\blue To start with,} we show in Fig.\ \ref{fig:16Pt2Y_OS_OC} the calculated layer-resolved  orbital ME susceptibilities $\bm{\chi}^L$ and orbital conductivities $\bm{\sigma}^{\bm L}$ for the $16$Pt/$2Y$ systems, for $\bm{M}\, || \,\bm{u}_z$, similar to the spin counterparts shown in 
Fig.\ \ref{fig:16Pt2Y_SS_SC}.
For the sake of completeness, we provide in Appendix \ref{App:orb-response} analogous plots to Fig.\ \ref{fig:NPt2Y_thickness_study} for the Pt-thickness dependence and to Figs.\ \ref{fig:Chi_S_inplane}, \ref{fig:Chi_S_outplane}, and \ref{fig:6Pt2Ni_SS_Mz_Mx_Vertical}  for the angular dependence  of the $\boldsymbol{\chi}^{L}$ tensors and their transformation properties under rotation of the magnetization direction. Also the dependence of the $\boldsymbol{\chi}^{L}$ tensor elements on the lifetime broadening is given in Appendix \ref{App:orb-response}. 

The layer-resolved results, shown in Fig.\ \ref{fig:16Pt2Y_OS_OC}(a), reveal that, just like for the spin, the $\bm{E}$-transverse component {\blue in the pure Pt system} resembles strongly the transport-induced accumulation of orbital angular momentum. {\blue The  transverse conductivity $\sigma_{zx}^{L_y}$ in Fig.\ \ref{fig:16Pt2Y_OS_OC}(d) is the orbital counterpart of the SHE, i.e., the OHE conductivity.}
 Notwithstanding the analogy to the spin response, the overall shapes of $\chi_{yx}^L$ and $\sigma^{L_y}_{zx}$ show distinct features when compared to their spin counterparts. The shape of the $\chi_{yx}^L$ profile is considerably less smooth {\blue and the flat area of $\sigma^{L_y}_{zx}$ in the interior of the Pt layer is far more extended {\blue for all systems.} In this area the orbital susceptibility, and thus the local accumulated orbital polarization, vanishes.}
 Notably, considering the values obtained, we obtain a huge orbital response $\chi_{yx}^L$, roughly one order of magnitude larger than the spin counterpart. {\blue Also the OHE conductivity (Fig.\ \ref{fig:16Pt2Y_OS_OC}(d)) is larger than the SHE conductivity.} This finding is consistent with previous calculations of the OHE in bulk metals, which obtained an intrinsic OHE that is much larger than the SHE
\cite{Tanaka2008,Kontani2009,Jo2018}.
{\blue The huge induced $E_{\bot}$-orbital susceptibility is consistent with previous calculations for noncentrosymmetric antiferromagnets that obtained an OREE that was much larger than the SREE \cite{Salemi2019}.}

When it comes to the relative magnitude of the different {\blue configurations, $\bm{E}_{\bot}$, $\bm{M}_{\bot}$, and $\bm{M}_{||}$,} striking differences compared to the spin responses can be observed. Here, the orbital response at the interface is dominated by the $\bm{E}$-transverse component. 
{\blue We associate the $\bm{E}$-transverse $\chi_{yx}^L$ component, as before, to the orbital accumulation caused by both the OREE and the OHE.} {\blue As we will see below, in particular the OREE (intraband) contribution to $\chi_{yx}^L$ is gigantic.}
The $\bm{M}$-transverse and $\bm{M}$-longitudinal orbital ME susceptibilities (Figs.\ \ref{fig:16Pt2Y_OS_OC}(b) and (c)), are an order of magnitude smaller.
Again, it is evident that the latter two orbital susceptibilities have a purely magnetic origin {\blue ($\bm{M}$-odd)} as they vanish for the nonmagnetic systems and are furthermore caused by the breaking of inversion symmetry. Similar to the case of the spin angular momentum, we identify the $\bm{M}$-transverse component $\chi_{xx}^L$ therefore as being an orbital polarization due to the {\blue \textit{magnetic}} OHE.

A further significant difference between the spin and orbital ME susceptibilities is the rapid variation of the orbital ME susceptibilities in the last few layers of the Pt/$Y$ interface. While the $\chi_{yx}^S$ 
 component has positive values for the {\blue atomic} monolayers in the vicinity of the interface (Fig.\ \ref{fig:16Pt2Y_SS_SC}(a)), the orbital counterpart exhibits a sign change for the two topmost layers.  A similar behavior can be observed for the $\bm{M}$-transverse components, $\chi_{xx}^S$ and $\chi_{xx}^L$. The unusual $\bm{M}$-longitudinal components exist, too, for the orbital ME susceptibility and conductivity,  Figs.\ \ref{fig:16Pt2Y_OS_OC}(c) and (f), but these quantities are, {\blue interestingly,} much smaller than their spin counterparts.

The dependence of the orbital responses on the Pt-layer thickness is shown in Fig.\ \ref{fig:NPt2Y_thickness_study_orb}
in Appendix A. Pt-layer thicknesses of about 8 monolayers provide stable values, for both  {\blue the $\bm{E}_{\bot}$ and $\bm{M}_{\bot}$ components} of the the orbital ME susceptibilities.

Orbital transport {\blue and orbital polarization} at interfaces are currently only poorly understood, and first measurements are being made \cite{Kim2020,Ding2020,Tazaki2020} as well as theory developed \cite{Go2020,Go2020b}.  Our calculations show that dependence of the orbital response $\boldsymbol{\chi}^L$ on the magnetization direction exhibits similarities with the spin response $\boldsymbol{\chi}^S$, as the nonzero components are the same for both cases. However, while the pair $\chi_{xy}^{S,\bm{u}_z}$/$\chi_{xy}^{S,\bm{u}_x}$ differs  close to the Pt/Ni interface, we find that $\chi_{xy}^{L,\bm{u}_z}$/$\chi_{xy}^{L,\bm{u}_x}$ are virtually identical. This suggests a different, much smaller, dependence of orbital
{\blue polarization} on the magnetization direction at an interface.

{\blue Next, we have investigated the  dependence of the induced orbital polarization on the magnetization direction. In the Appendix \ref{App:orb-response}, in Figs.\ \ref{fig:Chi_L_inplane} and \ref{fig:Chi_L_outplane}, we show the computed dependence of the components of $\bm{\chi}^L$ on the magnetization angles $\phi$ and $\theta$, respectively. The \textit{ab initio} computed  
 angle dependence of $\bm{\chi}^L$ bears several similarities to the angle dependence of the spin susceptibility $\bm{\chi}^S$, shown in Figs.\ \ref{fig:Chi_S_inplane} and \ref{fig:Chi_S_outplane}. The components of the tensor $\bm{\chi}^L$ follow a trigonometric dependence on the angles $\theta$ and $\phi$, but not identical to the ones given by Eqs.\ (\ref{chi-x-phi}) and (\ref{chi-z-theta}). However, for angles where a $\chi_{ij}^S$ component is zero, the corresponding $\chi_{ij}^L$ component is also zero. The symmetry of the $\bm{\chi}^L$ and $\bm{\chi}_{ij}^S$ tensor components with respect to $\bm{M}$, i.e., odd or even in $\bm{M}$, is also identical. A most significant difference is the dominance of the  $\bm{E}$-transverse components $\chi_{xy}^L
 \approx  -\chi_{yx}^L$ that are an order of magnitude larger than the other components.}

{\blue Further insight is obtained from considering the dependence of the orbital ME susceptibilities on the electronic lifetime. To exemplify the broadening effect we consider the 12Pt/2Co and 14Pt systems. In the Appendix B (Fig.\ \ref{fig:Broadening_Study_Orb}) we show the computed dependence of the nonzero orbital ME tensor elements on the lifetime broadening. 
Similar to the spin ME tensor elements, the intraband contributions increase linearly for decreasing broadening, whereas the interband contributions approach nonzero values (except for $\chi_{yx}^L$). The orbital ME susceptibilities reach however considerably larger values; for example, the intraband contribution to $\chi_{yx}^L$ is a factor of 150 larger than its spin counter part for the last Co atom of the 12Pt/2Co system.
This implies that in the limit of very pure crystals the $\bm{E}$-transverse interband $\chi_{yx}^L$ element, associated with the OREE, will dominate completely the response of this system. This conversely implies that the electrically-induced orbital polarization has then always a pure Rashba symmetry, i.e., the induced orbital moment is perpendicular to the electric field direction. This is in accordance with previous calculations for noncentrosymmetric antiferromagnets that showed that the OREE has perfect Rashba symmetry and is considerably larger than the SREE that does not give a perfectly $\bm{E}$-orthogonal spin response
\cite{Salemi2019}.}

\subsubsection{Dependence on spin-orbit coupling}
To investigate the dependence of the spin and orbital ME susceptibilities and conductivities we can vary the strength of the spin-orbit coupling in the calculations. To do this, we artificially introduce a SOC scaling parameter $\alpha$ in the DFT calculations such that $\hat{H}_0$ can be written as
$ \hat{H}_0 = \hat{H}_{\text{sc}} + \alpha \hat{H}_{\text{soc}}$ 
where $\hat{H}_{\text{sc}}$ is the scalar-relativistic part of the Hamiltonian and $\hat{H}_{\text{soc}}$ the SOC part. {\blue We find that without the $\hat{H}_{\text{soc}}$ term the whole spin ME susceptibility $\bm{\chi}^S$ and spin conductivity $\sigma_{ij}^{S_k}$, with indices such that $\epsilon_{ijk}= 0$, vanish.} {\blue Obviously, as the electric field $\bm{E}$ couples only to the electron's position operator $\hat{\bm{r}}$, relevant to the electron's orbital motion, SOC is necessary to provide a coupling to the spin.}
 Thus, these spin quantities are completely induced by the SOC. For $\boldsymbol{\chi}^L$, the story is quite different. When $\alpha$ is set to zero, $\chi^L_{xy}$ and $\chi^L_{xy}$, as well as $\sigma^{L_x}_{xz}$ and $\sigma^{L_x}_{yz}$, are nonzero and are actually not really affected by the modified SOC strength, a feature of the  that has been noted before for the OHE conductivities
\cite{Tanaka2008,Kontani2009} and for the
orbital polarizations induced by the OREE \cite{Salemi2019}. 

In Fig.\ \ref{fig:SOC_Scaling} we show comprehensive results for the layer-resolved profile of $\chi^S_{xy}$ and $\chi^L_{xy}$ for 6Pt/2Ni, computed for $\alpha = 0$, $0.1$, $0.5$, and $1$, with $\alpha = 1$ corresponding to the intrinsic SOC strength. It is evident from Fig.\ \ref{fig:SOC_Scaling}(a) that spin ME susceptibility is a pure SOC effect that scales linearly with the SOC. The situation is different for the orbital ME susceptibility, which exhibits practically no dependence on the SOC strength, see Fig.\ \ref{fig:SOC_Scaling}(b).  {\blue Clearly, the induced $\bm{E}$-transverse  orbital polarization represents the nonrelativistic response of the electron system to the electric field. That the induced orbital moment is nonrelativistic and perpendicular to the  electric field $\bm{E}$ can be recognized from the one-electron operator expression
\begin{equation}
\frac{d}{dt} (\delta \hat{\bm{L}}) = \hat{\bm{r}} \times e \bm{E}, 
\end{equation}
that does not require SOC.}
For all other spin and orbital susceptibility components,  as well as for the spin conductivity tensors elements, we find that these scale linearly with the size of the SOC, i.e., these are quantities induced by the SOC.

\begin{figure}[t!]
\centering
\includegraphics[width=0.85\linewidth]{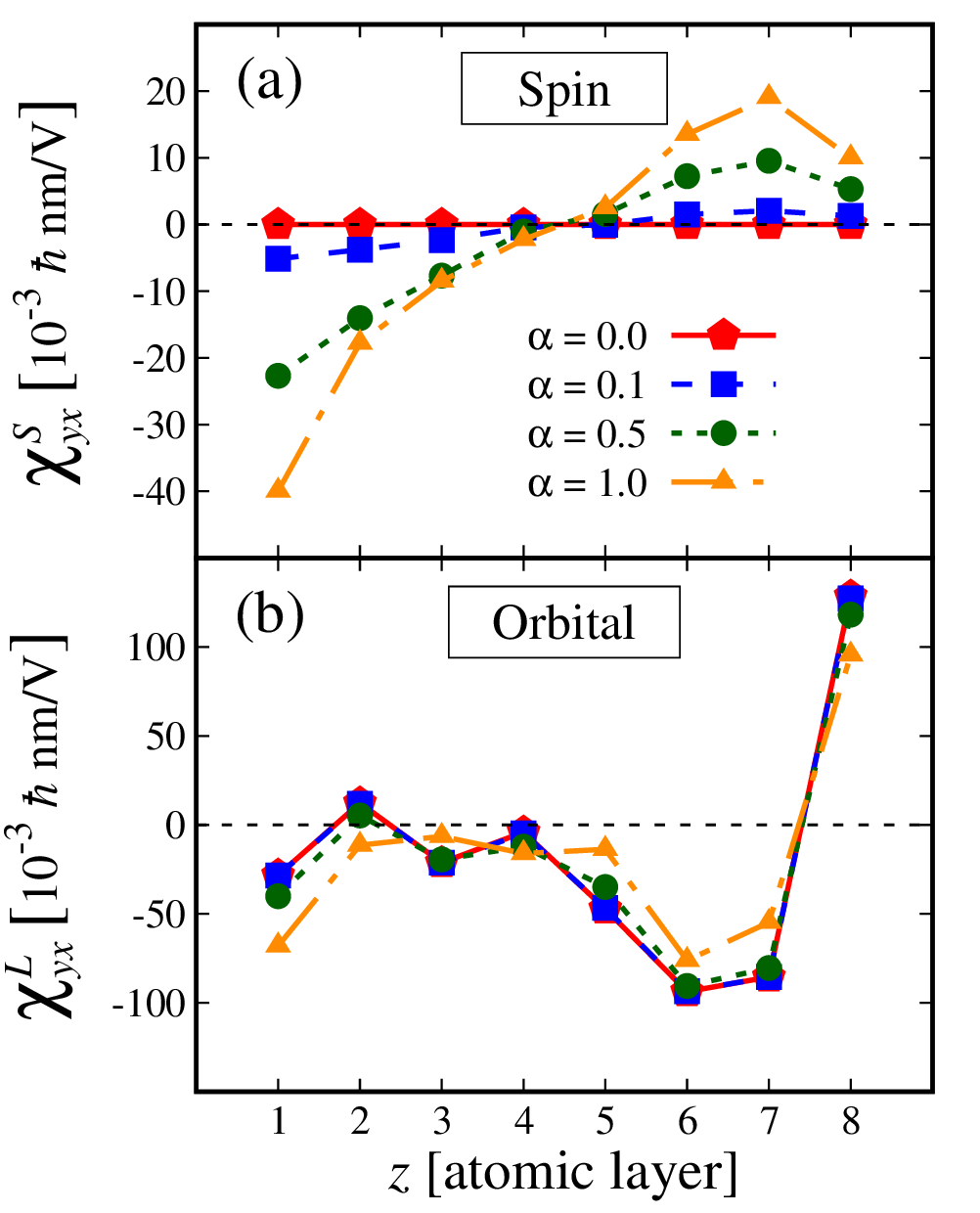}
\caption{Scaling behavior of (a) $\chi^S_{yx}$, and (b) $\chi^L_{yx}$ as a function of the SOC scaling parameter $\alpha$, calculated for the 6Pt/2Ni system ($\bm{M} \,|| \, \bm{u}_z$). 
 The $\bm{E}_{\bot}$ component of the spin ME susceptibility (a)  scales linearly with $\alpha$, and represents a SOC-induced quantity. 
 The $\bm{E}_{\bot}$ component of the orbital ME susceptibility (b) exists even without SOC.}

\label{fig:SOC_Scaling}
\end{figure}

\section{Discussion}
\subsection{Spin-orbit torque}
Freimuth \textit{et al.}\ \cite{Freimuth2015} evaluated directly the SOT using a different approach to the perturbative {\blue DFT} framework. While our computational method is distinct from theirs, we can evaluate the SOT  $\boldsymbol{\Tau}$ in a similar fashion. Using Eq.\ (\ref{eq:SOT-T-expression}), we can write
\begin{equation}
\delta \bm{B} \approx \underbrace{\frac{\langle V^{\downarrow}_{\text{KS}} - V^{\uparrow}_{\text{KS}}\rangle}{2\mu_B |\bm{S}|}
~ \boldsymbol{\Chi}^S}_{\boldsymbol{\Chi}_{\text{SOT}}} \, \bm{E},
\label{eq:BSOT}
\end{equation}
where $V^{\downarrow}_{\text{KS}}$ ($V^{\uparrow}_{\text{KS}}$) is the Kohn-Sham effective potential for minority (majority) spin electrons and $\bm{S}$ the equilibrium spin angular momentum. {\blue As mentioned in Sec.\ \ref{symmetry-SOT} this is an approximation of $\delta \bm{B}$.}

We define {\blue furthermore} $\boldsymbol{\chi}_{\textrm{\tiny{SOT}}}$ as the SOT spin susceptibility tensor in units of $\text{T} {\text{m}}{\text{V}}^{-1}$.  Since our computational approach involves quantities evaluated for each atomic site, we can access a layer-resolved $\bm{B}_{\text{SOT}}$.

For the thickest magnetic systems, $16$Pt/$2$Co and $16$Pt/$2$Ni, we find that the 
$\bm{E}_{\bot}$ contribution to the SOT at the first (second) layer of Ni is $ 0.0032$  $(0.0020)$ $ \text{mT}{\text{cm}}{\text{V}}^{-1}$ and $ 0.0019$  $ (0.0007) $ $ \text{mT}{\text{cm}}{\text{V}}^{-1}$ for Co. For the $\bm{M}_{\bot}$ contribution, we find $ 0.0020$  $(0.0030)$ $ \text{mT}{\text{cm}}{\text{V}}^{-1}$ for the first (second) layer of Ni and $ 0.0019$ $ (0.0020)$ $ \text{mT}{\text{cm}}{\text{V}}^{-1}$ the first (second) layer of Co. These values are smaller than, but consistent with, those obtained by Freimuth \textit{et al.}\ \cite{Freimuth2014}, because they used a much smaller broadening of electronic states. 

 {\blue The possible generation of large orbital torques has recently drawn attention \cite{Go2020}. However, although the orbital ME susceptibility is large, this does not automatically imply a large torque, because the induced orbital polarization can only couple to the static magnetic spin moment $\bm{M}$ via SOC. More precisely, in the commonly used formulation and implementation of DFT, 
the effective Kohn-Sham potential is not a functional of the orbital character, i.e., the exchange-correlation field only couples to spin angular momentum.}
{\blue The influence of the $\bm{\chi}^L$ on the SOT is nonetheless included in our calculations as the induced $\delta{\bm{L}}$ couples to ${\bm{S}}$ via SOC (and vice versa).}  Notwithstanding, theoretical efforts have recently been devoted to predicting the orbital torque \cite{Go2020,Go2020c} and experimental efforts are being undertaken to detecting the orbital polarization and torque and disentangling it from the spin torque \cite{Chen2018,Tazaki2020,Ding2020,Ding2021}.

\subsection{Relative sizes of {\blue fieldlike and dampinglike torques}}

\begin{figure}[t!]
\centering
\includegraphics[width=0.8\linewidth]{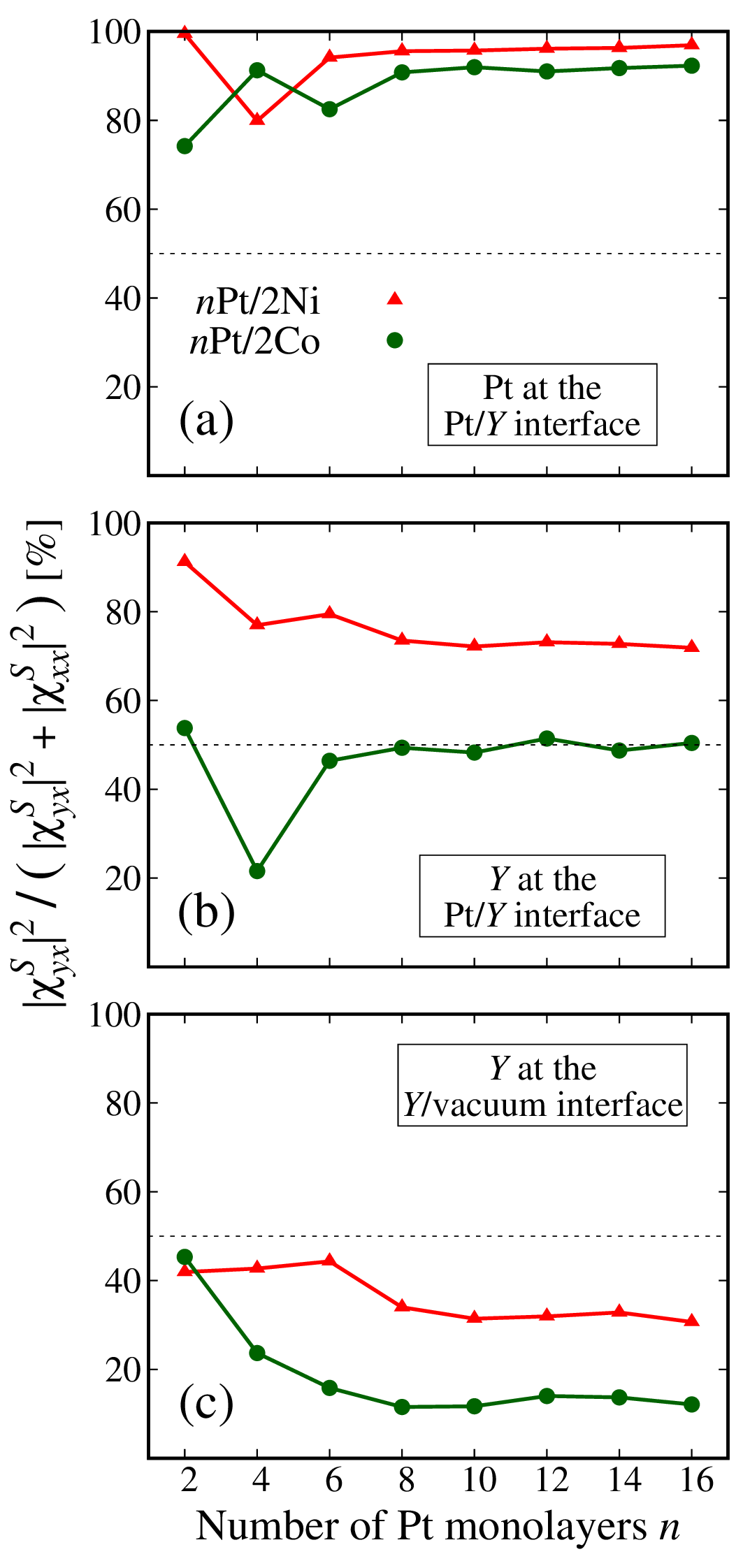}
\caption{Computed ratio of the fieldlike and (vectorial summed) total SOT as function of Pt layer thickness for the $n$Pt/2Co and $n$Pt/2Ni systems with $\bm{M}\,||\,\bm{u}_z$. The fieldlike $\bm{E}$-transverse torque is proportional to $\chi_{yx}^S$ and the dampinglike torque to $\chi_{xx}^S$. 
(a) The ratio at the Pt atom at the Pt/$Y$ interface, (b)  the ratio at the $Y$ atom at the Pt/$Y$ interface, and (c) at the $Y$ atom at the $Y$/vacuum interface ($Y$\,=\,Co or Ni).}
\label{fig:Transport_Vs_Local}
\end{figure}

The layer-resolved SOTs are dominantly defined by the size of the spin ME susceptibility $\chi^S$. Any resulting torque can be decomposed, as customary is done, in an even-in-$\bm{M}$ FL component and and odd-in-$\bm{M}$ DL component. The linear dependence of the effective SOT magnetic field on $\bm{\chi}^S$ permits to compare the relative sizes of the FL and DL SOTs. Considering the case $\bm{M}\, || \,\bm{u}_z$, the $\bm{E}_{\bot}$ component $\chi_{yx}^S$ leads to the FL torque, while the $\bm{M}_{\bot}$ component $\chi_{xx}^S$ corresponds to the DL torque. 
We can then quantify the relative importance of the two torques by computing the ratio
\begin{equation}
\frac{|\chi^{S}_{xy}|^2}{|\chi^{S}_{xy}|^2 + |\chi^{S}_{xx}|^2} \, \times \, 100 \% \, .
\end{equation}
A value of $ > 50 \%$ ($< 50 \%$) would then refer to a  {\blue fieldlike (dampinglike)} dominated   torque. The square exponent accounts for the fact that we are comparing vectorial quantities. Note that  $|\chi_{xx}^{S}|$ should be replaced by $|\chi_{zx}^{S}|$ for $\bm{M} \, || \, \,\bm{u}_x$.

The calculated Pt-thickness dependence of this ratio for the {\blue $n$Pt/2Ni and $n$Pt/2Co systems} is displayed in Fig.\ \ref{fig:Transport_Vs_Local}, {\blue for the last Pt monoatomic layer at the Pt/$Y$ interface, as well as for the $Y$ monolayer at the Pt interface and at the vacuum interface.} {\blue It can be observed that} there is virtually no change for the computed torque ratio for Pt layer thicknesses beyond eight Pt monolayers. 
%The  ratio reaches a stable value as function of the Pt-layer thicknesses for eight Pt monolayers. 
For the Pt monolayer at the Pt/$Y$ interface, the induced torque is to $90 \%$ composed of the {\blue FL torque} component,  see Fig.\ \ref{fig:Transport_Vs_Local}(a). For the $Y$ monolayer at the Pt/$Y$ interface, the torque consists for $\sim 75 \%$ of the {\blue FL} component for $Y=$ Ni and $\sim 50 \%$ for $Y=$ Co (Fig.\ \ref{fig:Transport_Vs_Local}(b)). For the $Y$ monolayer at the $Y$/vacuum interface, the torque consist for  $\sim 30 \%$ of the {\blue FL} component for $Y=$ Ni and $\sim 10 \%$ for $Y=$ Co (Fig.\ \ref{fig:Transport_Vs_Local}(c)). This suggests that the Pt/Ni interface is more transparent to spin currents from the Pt than the Pt/Co interface, 
consistent with the better matching electronic structures of isoelectronic fcc Ni and Pt.

The torques resulting from the induced spin polarization on the two ferromagnetic $Y$ monolayers will be the most important ones for the magnetization switching. The torque on the ferromagnetic layer at the vacuum interface is thus {\blue approximately both field- and dampinglike}, whereas the torque at the ferromagnetic layer adjacent to the Pt layer has a larger {\blue FL} contribution. {As the relative contribution of the $\bm{E}_{\bot}$ and $\bm{M}_{\bot}$ components differs in both $3d$ monolayers},
the direction of the total torque per monolayer will be different for each of the two $Y$ monolayers. 
The calculated atomic-layer specific torques are ideally suited to investigate current-driven magnetization switching dynamics using atom-specific Landau-Lifshitz-Gilbert spin-dynamics simulations (see e.g.\ \cite{Nowak2007,Evans2014,Jungfleisch2017}). Such simulations would provide {\blue layer-specific} insight in how the magnetization of the ferromagnetic layers reverses {\blue under an applied electric field.}

%%%%%%%%%%%%%%%%%%%%%%%%%%
%%% CONCLUSION SECTION %%%
%%%%%%%%%%%%%%%%%%%%%%%%%%
\section{Conclusions}
We have employed first-principles calculations to investigate
the electric-field induced spin and orbital magneto-electric susceptibility and the spin and orbital conductivity of heavy-metal/$3d$-metal bilayer structures. For each orientation of the $3d$ magnetization and the applied electric field we have shown that the susceptibility tensor and its associated conductivity tensor can be uniquely decomposed in components depending on the spatial symmetries, i.e., 
transverse electric $\bm{E}_{\bot}$, transverse magnetic $\bm{M}_{\bot}$, and longitudinal magnetic components $\bm{M}_{\parallel}$, as well as the  magnetic symmetries (odd-in-$\bm{M}$ and even-in-$\bm{M}$, respectively). Our atomic-layer specific calculations of the tensors show that all components are highly dependent on the position of the atomic layer in the considered heterostructure.

Analyzing the properties of the computed ME susceptibilities, we have identified the even-in-$\bm{M}$, $\bm{E}_{\bot}$-components of $\bm{\chi}^S$ as spin accumulation associated with both the SHE and SREE, and the odd-in-$\bm{M}$, $\bm{M}_{\bot}$-components  with the magnetic SHE. {\blue We note however that our \textit{ab initio} formulation uses a more general form of the SOC than the often used more elementary Rashba SOC.}
Extending the calculations to field-induced orbital polarizations, we have performed a similar analysis and decomposition for the orbital susceptibility tensor $\bm{\chi}^L$ and orbital conductivity, $\bm{\sigma}^L$.
We have analyzed the relative importance of the different spin and orbital contributions as a function of Pt thickness. Both the out-of-equilibrium  $\bm{E}_{\bot}$ and $\bm{M}_{\bot}$ spin responses lead to atomic-layer dependent SOTs that are of the same order of magnitude, but act in perpendicular directions. 
We find that the $\bm{E}$-transverse spin accumulation is largest for the Pt layer at the Pt/$3d$-metal interface. The $\bm{M}$-transverse spin accumulation, {\blue conversely,} is larger at the $3d$-vacuum interface. Our calculations show that both effects should be considered \textit{together} when analyzing current-induced spin polarization in  heavy-metal/ferromagnetic bilayer systems.

{\blue This perception is valid for electronic relaxation times  that are realistic for metallic systems ($\hbar/\tau \approx 0.25$ eV). For extremely pure materials, however, the intraband contribution to the $\bm{E}$-transverse induced spin polarization will  be much larger than other (interband) contributions, leading to a predominant fieldlike SOT.}

{\blue Considering} the electric-field induced orbital polarization, we find that the $\bm{E}$-transverse orbital susceptibility and conductivity components are always much larger ($\sim 10 \times$) than {\blue their $\bm{M}$-transverse orbital counterparts.} {\blue In contrast to the spin counterparts, the $\bm{E}_{\bot}$-orbital ME susceptibility and OHE conductivity do not} 
dependent on SOC. {\blue This exemplifies that the induced $\bm{E}$-transverse orbital polarization is the primary response of the electron system to the electric field and that the other, both spin and orbital, induced polarizations are generated from the nonzero $\bm{E}_{\bot}$-orbital susceptibility by SOC.} {\blue The nonrelativistic orbital ME susceptibilities are  ten to a hundred times larger than the corresponding relativistic spin susceptibilities.} However, although the {\blue electrically} induced orbital polarization is huge, it can only couple to the equilibrium spin moment via SOC.

{\blue The computed induced spin and orbital polarizations follow trigonometric functional dependencies on the magnetization direction angles that are consistent with the $\bm{M}$ symmetry of the ME susceptibility components.  Of particular interest are the large $\bm{E}_{\bot}$-orbital ME components that are practically angle independent and antisymmetric, i.e.,
$\chi_{xy}^L = - \chi_{yx}^L$.  The induced $\bm{E}$-transverse orbital polarization exhibits consequently a {pure} Rashba symmetry.
} 

Our calculations show furthermore that there exists as well an electric-field induced spin and orbital polarization along the magnetization direction. This previously unobserved spin-orbit effect  does not exert a torque on the static magnetization. We propose that it could be possible to observe this $\bm{M}$-longitudinal effect in sensitive magneto-optical Kerr effect measurements (cf.\ \cite{Stamm2017}).

When the magnetization direction changes, the spin and orbital responses also change. We have shown that the magnetization direction does have a strong influence on the spin and orbital responses, but that it is possible to track the evolution of the individual components using simple, but robust, symmetry relations. This should aid the {\blue future} investigation of SOT magnetization switching using atom-specific Landau-Lifshitz-Gilbert spin-dynamics simulations.

\begin{acknowledgments}
We thank the anonymous reviewer for constructive comments. This work has been supported by the Swedish Research Council (VR), the K.\ and A.\ Wallenberg Foundation (Grant No.\ 2015.0060), the European Union's Horizon2020 Research and Innovation Programme (Grant agreement No.\ 863155, s-Nebula), and the Swedish National Infrastructure for Computing (SNIC). The calculations were performed at the PDC Center for High Performance Computing and the Uppsala Multidisciplinary Center for Advanced Computational Science (UPPMAX).
\end{acknowledgments}

\begin{appendix}
\section{Computational details}
\label{ap:Computational_Details}
As mentioned in Sec.\ \ref{CompMeth}, the calculations are performed in 3 steps. First, the structures are fully relaxed with the DFT package SIESTA \cite{Soler2002}. The cell parameters and atomic positions of the pure Pt films are relaxed until the pressure reaches values below 0.001 GPa and atomic forces on each atom are below 0.01 eV/{\AA}. Then, the cell parameter is fixed and two monolayers of 3$d$ elements (Ni, Co, or Cu) are added. The atomic positions are then relaxed using the same criterion as before. All SIESTA calculations are performed using a $15 \times 15 \times 1$ Monkhorst-Pack grid \cite{Monkhorst1976} with an electronic temperature of 300\,K. The double $\zeta$ with polarization pseudo-atomic basis set functions are used. The mesh-cutoff for real space integration is set to 250\,Ry and we use the generalized gradient approximation (GGA) for the exchange-correlation functional in the PBEsol parametrization \cite{Perdew2008}. All structures contain $20$ {\AA} of vacuum to avoid spurious interactions with neighboring simulation cells.

Second, once the structures are relaxed, the ground-state Kohn-Sham wavefunctions and energies are computed using the accurate full-potential, all-electron code WIEN2k \cite{Blaha2018}, with spin-orbit interaction included \cite{Kunes2001}. The product between the smallest muffin-tin radius $R_{MT}$ and the largest reciprocal vector $K_{max}$ is set to $R_{MT} \times K_{max} = 8.5 $ and the self-consistent spin-polarized density is computed using a $30\times 30 \times 1$ $k$-points Monkhorst-Pack grid. The computed spin moments for the 16Pt/2Ni bilayer are 0.855 $\mu_B$ and 0.760 $\mu_B$ at Ni18 and Ni17, respectively. The spin moment on the Pt interface layer (Pt16) is 0.212 $\mu_B$. For the 16Pt/2Co bilayer the equivalent moments are 2.02 $\mu_B$, 1.942 $\mu_B$, and 0.251 $\mu_B$. The proximity induced moments in the Pt layer vanish within four layers.

Finally, the atom-resolved spin response tensors are then computed with a denser $200 \times 200 \times 1$ k-mesh. As the WIEN2k code uses atom-centered wavefunctions, we exploit here this property to compute them in an atom-projected fashion. The simulation cell is divided into two subspaces: muffin-tin spheres around each atom, in which the wavefunction is expanded in terms of spherical harmonics, and the interstitial region in which the wavefunction is given in terms of plane waves, i.e.,
\begin{equation}
\Psi(\bm{r}) = \sum_\alpha \Psi_\alpha(\bm{r}-\bm{R}_\alpha) + \Psi_I (\bm{r}),
\end{equation}
where the first right-hand term is the wavefunction about atom $\alpha$ and $\Psi_I(\bm{r})$ is the wavefunction in the interstitial. The atom-projected expected value of an operator $\hat{O}$ is taken as
\begin{equation}
O_\alpha = \int d {\bm{r}} \, \Psi_\alpha^*(\bm{r}-\bm{R}_\alpha) \hat{O} \Psi_\alpha(\bm{r}-\bm{R}_\alpha) ,
\end{equation}
where the integral is over the $\alpha^{\rm th}$-muffin-tin volume and $\hat{O}$ can be replaced by any operator described in the Sec.\ \ref{secIIB}.

\section{Properties of the orbital responses}
\label{App:orb-response}

In this Appendix, we provide detailed calculated results for the orbital susceptibilities. 
We show in Fig.\ \ref{fig:NPt2Y_thickness_study_orb} 
the calculated dependence of the layer-resolved $\bm{\chi}^L$ tensor elements on the number of Pt monolayers $n$,
similar to the {\blue results} for the spin counterpart shown in 
Fig.\ \ref{fig:NPt2Y_thickness_study}.
In Figs.\ \ref{fig:Chi_L_inplane} and \ref{fig:Chi_L_outplane} we provide the selfconsistently calculated dependence of the tensor elements on the magnetization angles $\phi$ and $\theta$, respectively, for the 12Pt/2Co system. The angular dependencies are similar to the ones computed for the induced spin susceptibilities, shown in Figs.\ \ref{fig:Chi_S_inplane} and
\ref{fig:Chi_S_outplane}, and are even or odd-in-$\bm{M}$, consistent with the symmetry classification. The slight asymmetry of some tensor elements, e.g., $\chi_{xx}^L$ and $\chi_{yy}^L$ 
in Fig.\ \ref{fig:Chi_L_inplane}, can be ascribed to an angular dependence of the form $\sin 2\phi / (a \cos^2 \phi + b \sin^2 \phi)$, with nonequal constants $a$ and $b$.

\newpage

In Fig.\ \ref{fig:6Pt2Ni_OS_Mz_Mx_Vertical} 
 we provide the computed magnetization-direction dependence  of the $\boldsymbol{\chi}^{L}$ for magnetization directions $\bm{M} \, || \, \bm{u}_z$ to $\bm{M} \, || \, \bm{u}_x$. The similar shape of the curves and the very similar values of the $\bm{\chi}^L$ tensor elements illustrates the mapping according to the employed classification.
In Fig.\ \ref{fig:Broadening_Study_Orb}, lastly, we show the computed dependence of the intraband and interband orbital susceptibility elements on the electronic broadening, for the 12Pt/2Co and 14Pt systems.

\mbox{}
\newpage

\begin{figure*}[tbh!]
\centering
\includegraphics[width=0.85\linewidth]{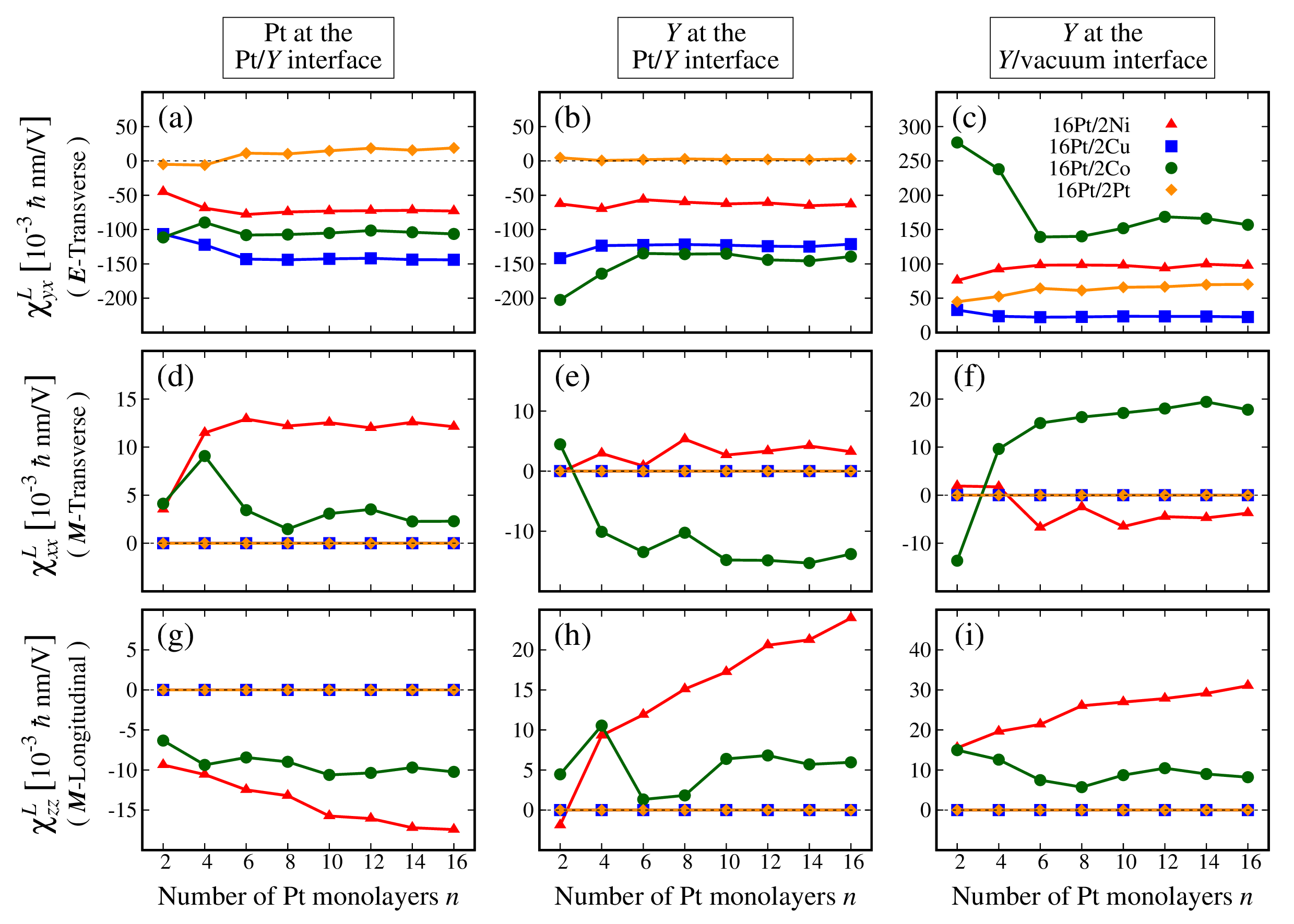}
\caption{Dependence of the orbital susceptibility tensor $\boldsymbol{\chi}^L$ on the number of Pt monolayers $n$. Top row:  $\bm{E}$-transverse component of $\boldsymbol{\chi}^L$ for (a) the Pt atom at the Pt/$Y$ interface, (b) the $Y$ atom at the Pt/$Y$ interface, and (c) the $Y$ atom at the $Y$/vacuum interface. Middle row: $\bm{M}$-transverse component of $\boldsymbol{\chi}^L$ for (d) the Pt atom at the Pt/$Y$ interface, (e) the $Y$ atom at the Pt/$Y$ interface, and (f) the $Y$ atom at the $Y$/vacuum interface. Bottom row: $\bm{M}$-longitudinal component of $\boldsymbol{\chi}^L$ for (g) the Pt atom at the Pt/$Y$ interface, (h) the $Y$ atom at the Pt/$Y$ interface, and (i) the $Y$ atom at the $Y$/vacuum interface.}
\label{fig:NPt2Y_thickness_study_orb}
\end{figure*}
%\mbox{}

%\nopagebreak

%\FloatBarrier

\begin{figure}[bht!]
\centering
\includegraphics[width=0.98\linewidth]{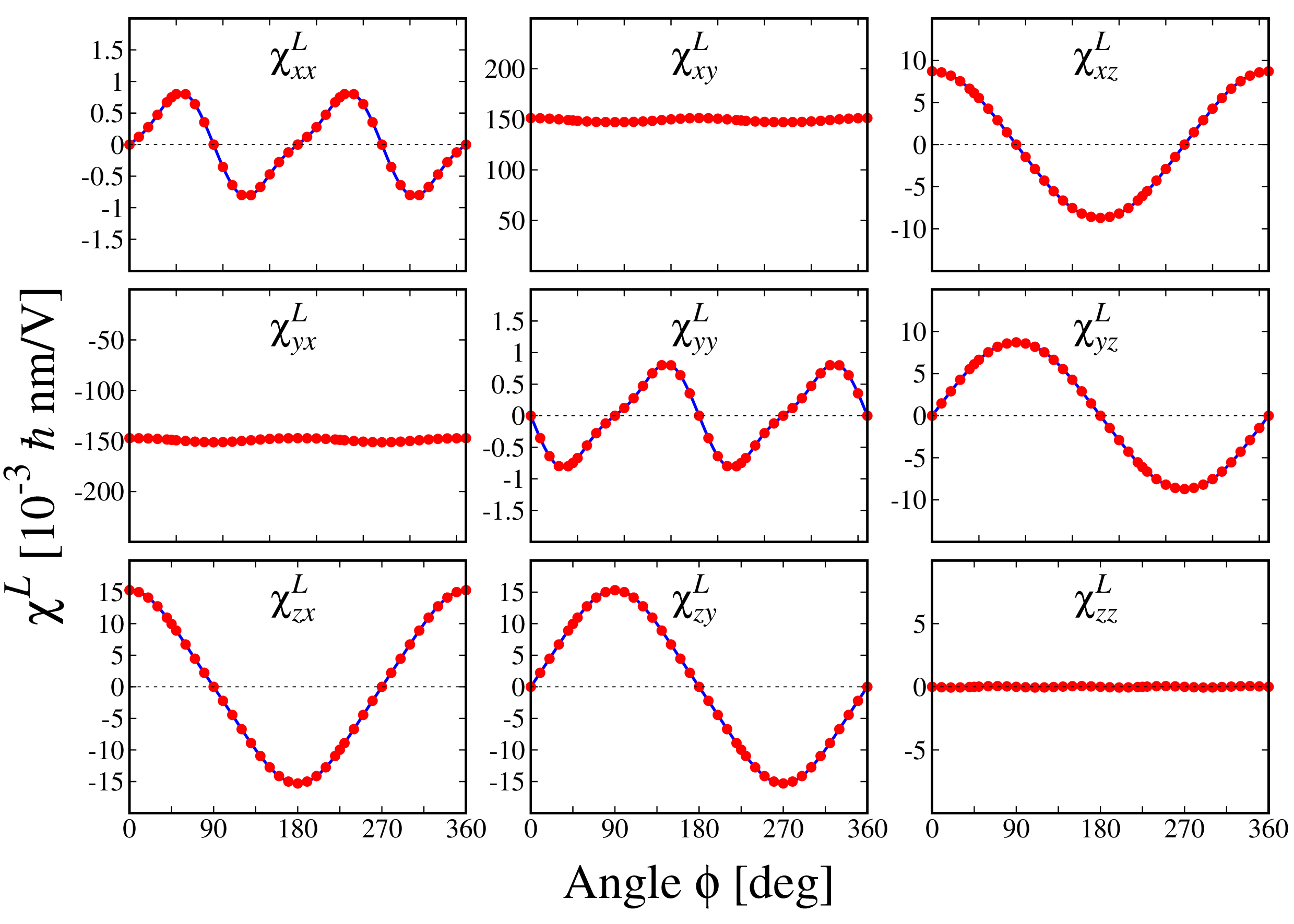}
\caption{Computed angular dependence of the orbital susceptibility tensor $\bm{\chi}^L$ of the 12Pt/2Co system at the interfacial Co atom for in-plane magnetizations, given as a function of the azimuthal angle $\phi$.}
\label{fig:Chi_L_inplane}
\end{figure}

\begin{figure}[bht!]
\centering
\includegraphics[width=0.98\linewidth]{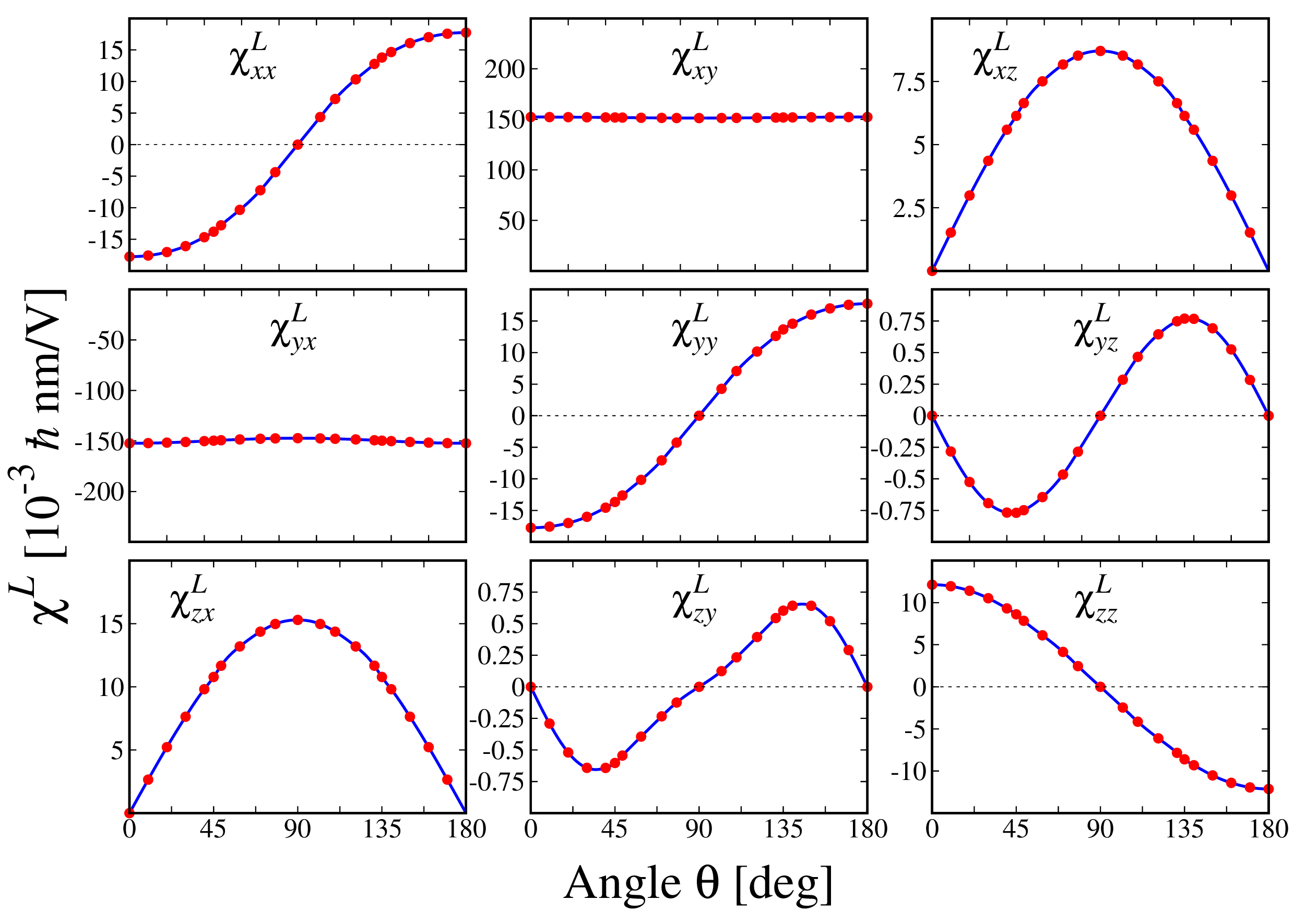}
\caption{Computed angular dependence of the orbital susceptibility tensor $\bm{\chi}^L$ of the 12Pt/2Co system at the interfacial Co atom for magnetizations in the $x - z$ plane, given as a function of the polar angle $\theta$.}
\label{fig:Chi_L_outplane}
\end{figure}

\mbox{}
\newpage

\begin{figure}[ht!]
\centering
\includegraphics[width=0.80\linewidth]{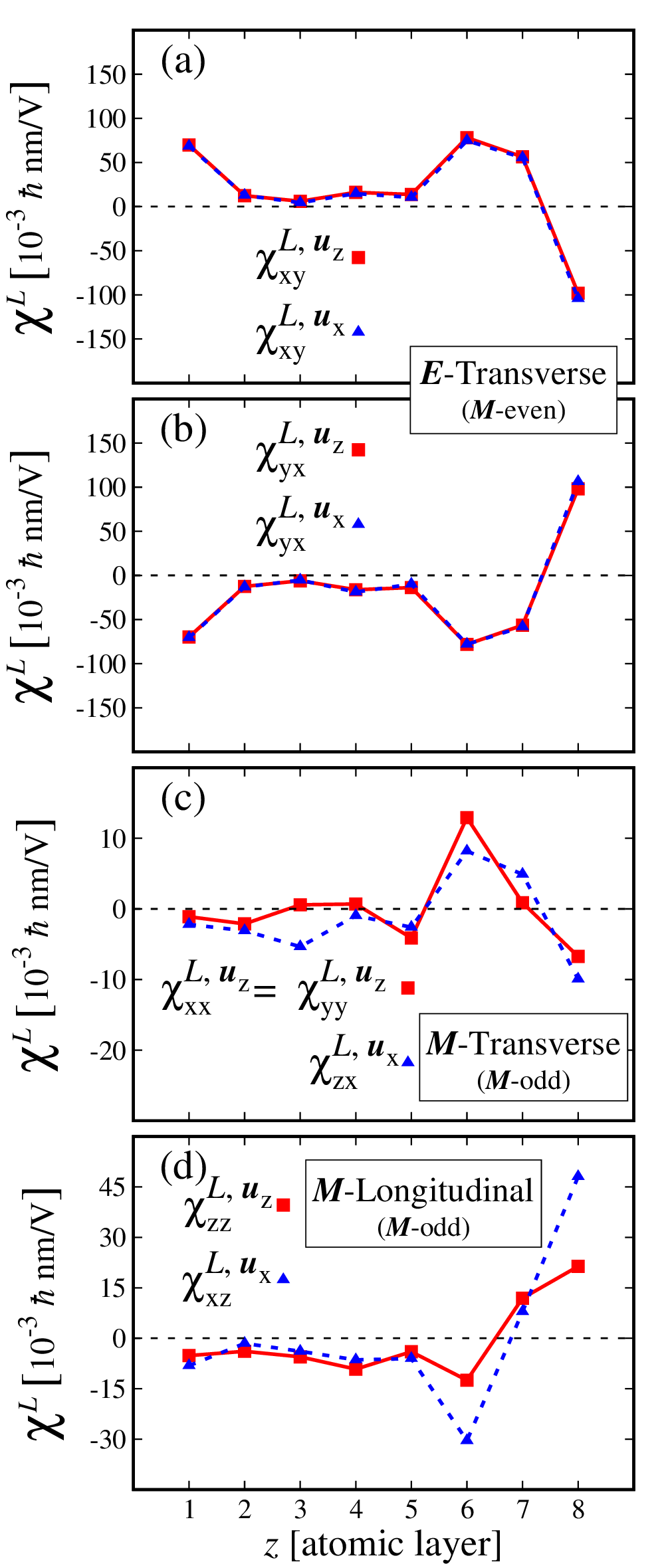}
\caption{Magnetization direction dependence of the orbital susceptibility tensor $\boldsymbol{\chi}^L$ for $6$Pt/$2$Ni. The data are shown for $\bm{M} \, || \, \bm{u}_z$ ($\bm{M} \, || \, \bm{u}_x$) with the red squares (blue triangles). When the magnetization direction switches from $\bm{u}_z$ to $\bm{u}_x$, the transverse components $S_{xy}^{L,\bm{u}_z}$ and $S_{yx}^{L,\bm{u}_x}$ are mapped onto themselves.   The $\bm{M}$-transverse components $\chi_{xx/yy}^{L,\bm{u}_z}$ are mapped onto $\chi_{zx}^{L,\bm{u}_x}$.  The $\bm{M}$-longitudinal component $\chi_{zz}^{L,\bm{u}_x}$  is mapped onto $\chi_{xz}^{L,\bm{u}_x}$. The mapping of the components upon magnetization rotation respects our symmetry analysis, in terms of $\bm{E}_{\bot}$, $\bm{M}_{\bot}$, and $\bm{M}_{||}$ principal components.}
\label{fig:6Pt2Ni_OS_Mz_Mx_Vertical}
\end{figure}

\mbox{}
\newpage

\begin{figure}[ht!]
\centering
\includegraphics[width=0.70\linewidth]{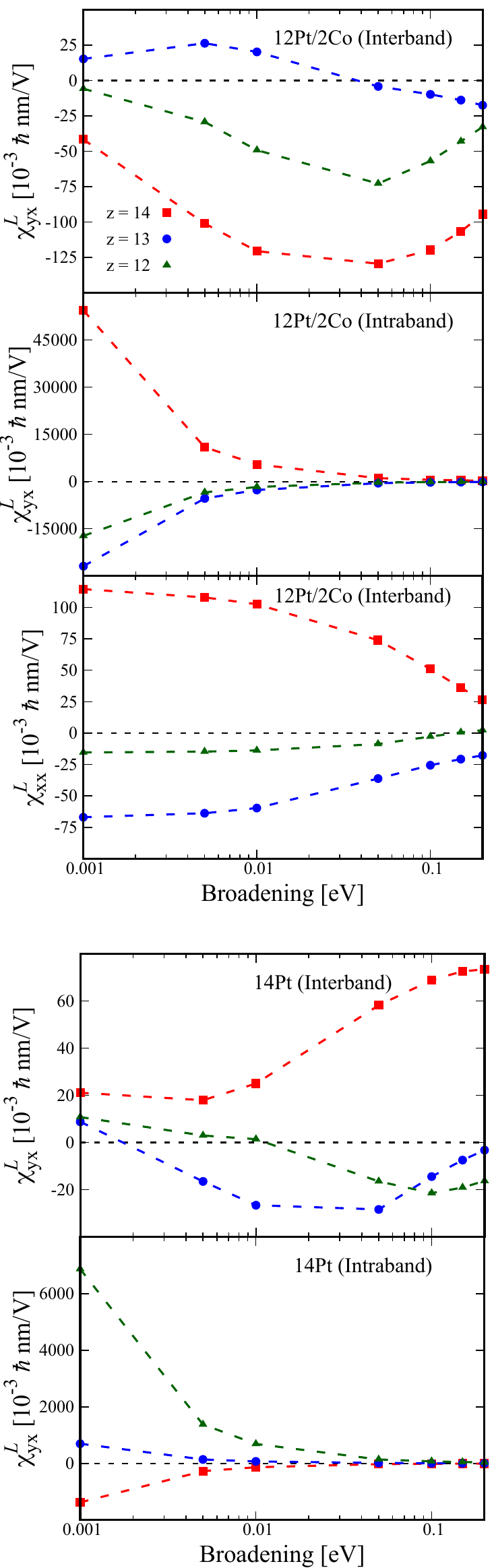}
\caption{Dependence of the interband and intraband contributions of the orbital ME response $\chi^L$ on the electronic broadening $\delta$ for the 12Pt/2Co and 14Pt systems. 
%The $xx$ component is purely interband while $xy$ contains both interband and intraband. Note that due to the purely quantum nature of the $xx$ component semi-classical approaches such as Boltzmann transport theory fail at capturing this component. The broadening used in this paper, $\delta = 0.2$ eV is a realistic value for metallic systems.
Note that the orbital ME response is roughly one to two orders of magnitude larger than the spin ME response. }
\label{fig:Broadening_Study_Orb}
\end{figure}

\mbox{}
\newpage

\end{appendix}

%\mbox {}
\newpage

\mbox {}
\newpage
\end{document}